%% file: paper.tex
\documentclass[10pt,sigconf,letterpaper,nonacm]{acmart}
\renewcommand\footnotetextcopyrightpermission[1]{}
\usepackage{booktabs} 
\usepackage{subcaption}
\usepackage{enumerate}
\usepackage{cleveref}
\crefformat{footnote}{#2\footnotemark[#1]#3}

\setcopyright{none}

\usepackage{multirow}

\usepackage{listings}

\usepackage{xcolor}
\definecolor{dark green}{RGB}{94,145,40}

\providecommand{\ione}{\emph{(i)} }
\providecommand{\itwo}{\emph{(ii)} }
\providecommand{\ithree}{\emph{(iii)} }

\providecommand{\ie}{\emph{i.e.,} }
\providecommand{\eg}{\emph{e.g.,} }

\providecommand{\etc}{\emph{etc.}}

\usepackage{listings}
\usepackage{xcolor}
\colorlet{punct}{red!60!black}
\definecolor{background}{HTML}{EEEEEE}
\definecolor{delim}{RGB}{20,105,176}
\colorlet{numb}{magenta!60!black}

\lstdefinelanguage{json}{
    basicstyle=\normalfont\ttfamily,
    numbers=left,
    numberstyle=\scriptsize,
    stepnumber=1,
    numbersep=4pt,
    showstringspaces=false,
    breaklines=true,
    frame=lines,
    backgroundcolor=\color{background},
    literate=
     *{0}{{{\color{numb}0}}}{1}
      {1}{{{\color{numb}1}}}{1}
      {2}{{{\color{numb}2}}}{1}
      {3}{{{\color{numb}3}}}{1}
      {4}{{{\color{numb}4}}}{1}
      {5}{{{\color{numb}5}}}{1}
      {6}{{{\color{numb}6}}}{1}
      {7}{{{\color{numb}7}}}{1}
      {8}{{{\color{numb}8}}}{1}
      {9}{{{\color{numb}9}}}{1}
      {:}{{{\color{punct}{:}}}}{1}
      {,}{{{\color{punct}{,}}}}{1}
      {\{}{{{\color{delim}{\{}}}}{1}
      {\}}{{{\color{delim}{\}}}}}{1}
      {[}{{{\color{delim}{[}}}}{1}
      {]}{{{\color{delim}{]}}}}{1}
      {\ \ }{{\ }}1,
}

\usepackage{pifont}
\newenvironment{itemize-s}%
{\begin{itemize}%
		\setlength{\itemsep}{0pt}%
		\setlength{\parskip}{0pt}}%
	{\end{itemize}}

\usepackage{multirow}

\begin{document}
\title[A hop away from everywhere]{A hop away from everywhere: A view of the intercontinental long-haul infrastructure}


\author{Esteban Carisimo}
\affiliation{%
  \institution{Northwestern University}
}

\author{Mia Weaver}
\affiliation{%
  \institution{University of Wisconsin-Madison}
}

\author{Paul Barford}
\affiliation{%
  \institution{University of Wisconsin-Madison}
}

\author{Fabi\'an E. Bustamante}
\affiliation{%
  \institution{Northwestern University}
}

\renewcommand{\shortauthors}{Carisimo et al.}

\begin{abstract}
\input{tex/abstract.tex}
\end{abstract}

\settopmatter{printfolios=true}
\maketitle

\begin{abstract}
\input{tex/abstract.tex}
\end{abstract}

\input{tex/intro.tex}
\input{tex/methodology.tex}
\input{tex/dataset.tex}
\input{tex/observations.tex}

\input{tex/longitudinal.tex}
\input{tex/implications.tex}
\input{tex/discussion.tex}
\input{tex/relwork.tex}
\input{tex/conclusions.tex}


\bibliographystyle{ACM-Reference-Format} 
\bibliography{ref}

\appendix\newpage
\input{tex/appendix.tex}

\end{document}

%% file: tex/abstract.tex
Over the past two decades, a desire to reduce transit cost, improve control over routing and performance, and enhance the quality of experience for users, has yielded a more densely connected, flat network with fewer hops between sources and destinations.  The shortening of paths in terms of the number of hops or links has also meant, for what is at the end an infrastructure-bound network, the {\em lengthening} of many of these links.  
In this paper, we focus on an important aspect of the evolving logical connectivity of the Internet that has received little attention to date: intercontinental long-haul links.
We develop a methodology and associated processing system for identifying long haul links in traceroute measurements.  We apply this system to a large corpus of traceroute data and report on multiple aspects of long haul connectivity including country-level prevalence, routers as international gateways, preferred long-haul destinations, and the evolution of these characteristics over a 7 year period.  
We identify over 9K layer 3 links that satisfy our definition for intercontinental long haul with many of them terminating in a relatively small number of nodes.  
An analysis of connected components shows a clearly dominant one with a relative size that remains stable despite a significant growth of the long-haul infrastructure.

%% file: tex/intro.tex
\section{Introduction}
\label{sec:intro}

The consolidation of the Internet as a multimedia network has brought fundamental changes
to its topology. A desire to reduce transit cost, improve control over routing and performance, 
and enhance the experience for users, have yielded a more densely connected, flat network of shorter 
paths~\cite{labovitz2010internet, dhamdhere2010internet}. Understanding these trends, the forces 
behind them, and their implications have attracted significant research attention~\cite{carisimo2019studying, 
chiu2015we, bottger2018looking, gigis2021seven, calder2013mapping, bottger2018open, hayes2008cloud,
greengard2010cloud, ager2012anatomy,carisimo2020first, fanou2015diversity, gupta2014peering}. In 
this evolving view of the Internet, Content Delivery Networks (CDNs), cloud computing, and IXPs 
have displaced Tier-1 transit providers from center stage, suggesting a diminished role for long-haul 
connectivity.

Anecdotal evidence and some recent studies, however, suggest that the long-haul infrastructure -- 
both terrestrial and submarine cables -- still plays a critical role in the Internet ecosystem. 
Recurrent cable cuts and failures show that, despite there being multiple cables between continents, a single 
cable cut can result in large outages or a country-wide Internet blackout~\cite{ar2021outage, vn2021outage, 
co2021outage, mr2018outage, ar2017outage}. Even popular regional websites for landlocked countries 
rely on objects hosted in locations only reachable via submarine long-haul links~\cite{liu2020out}. 

The focus of this paper is the intercontinental long-haul infrastructure. We take a network-layer 
perspective to investigate this infrastructure, developing a methodology and associated processing system 
for identifying long-haul links (LHLs) in large traceroute datasets, their preferred destinations, and the 
resulting long-haul link network. 

We apply this methodology to a large corpus of traceroute data (231.45M traceroutes), collected by the 
CAIDA Archipelago platform, and report on multiple aspects this network. Focusing on a recent snapshot 
(composed of three measurement \textit{cycles}\footnote{Each Ark cycle is a traceroute campaign covering 
all /24 subnets from a single vantage point~\cite{caida-probing-ipv4}.}), we identified 31,452 LHLs directly 
connecting 15,031 routers in nearly every country of the world (146 or $\approx$79\% of all countries), with 
10\% of LHLs connecting in a single hop routers 193ms away in countries separated by over 12,500km. 
Such LHLs may hide important details about the underlying physical paths, potentially relevant 
to cyber-sovereignty discussions~\cite{10.1145/2785956.2787509, obar2012internet, hathaway2014connected}, and challenge existing
approaches from root cause analysis of failures to congestion control.


We discovered preferred destinations in the LHnet where the 80.1\% of the LHL terminate in the US, and 66.9\% 
interconnect nodes in the North Atlantic. We explore properties of these node and identify 
high-degree vertices, or {\em super routers}, connecting up to one-fourth of all countries in our dataset 
(28 countries), with most of them operated by TIER-1 transit providers and $\approx$76\% (352/462) of them located 
in the US. The existence of a few well-connected routers raises concerns about security and resilience~\cite{doi:10.1073/pnas.172501399}. Given the distribution of LHLs distances, it is possible that these 
LHLs are the network-level view of long-haul link infrastructure, such as submarine cables, and that popular 
destinations for these LHLs are at or nearby landing points in coastal areas. Our analysis, however, identify a 
number of these high-degree vertices far in-land, even as far as Chicago (US). 


Many of the identified LHLs result from the wide adoption of 
virtualization technologies (e.g., MLPS), which hide physical links in virtual network-layer link connecting 
pair of nodes, as far as Sao Paulo and Tokyo, to each other in a single hop. We investigated the adoption of 
identifiable MPLS tunnels and found heterogeneous adoption across networks,  7.79\% on average but with prominent 
examples of large adoption such as Vodafone-AS1273 using in it in 98.2\% of its LHLs. 
The adoption of MPLS tunnels challenges routing optimization and debugging in presence of path inflation~\cite{bozkurt2018dissecting, krishnan2009moving}, in some cases resulting from {\em autobandwidth} algorithms~\cite{pathak2011latency}.


We carry out a longitudinal study of the LHL graph, we explore topological changes in the LHL graph over a 
multi-year period starting in 2016. In just seven years, the number of edges grew 2.2x, from 7,857 to 17,224,  
while the number of vertices doubled (to 9,802). Despite this growth, some properties of the LHnet has remained stable. 
In this 7-year period the inter-hop latency distribution has remained the same and the prevalence of intra-AS LHLs has had minor changes (72\% to 87\%). 

In sum, we make the following key contributions: 
\begin{itemize}
	\item We develop a new methodology that identifies LHLs from traceroute measurements.
	\item We apply the methodology to a large corpus of traceroute data collected world-wide over a period of 7 years.
	\item We find that LHL are a significant and growing feature in today's layer 3 connectivity, and that LHLs predominately terminate in super routers, the majority of them found in the United States.
\end{itemize}


We plan to release source code and artifacts to facilitate reproducibility of our study.



%% file: tex/methodology.tex
\section{Long-haul links}
\label{sec:definition}

We focus our study on the intercontinental long-haul infrastructure from a 
network-layer view. Key to our analysis is the identification of long-haul 
links in large traceroute datasets. In the following paragraphs, we present a working 
definition of these links before discussing our inference methodology.

\subsection{A working definition}
 
We define a {\em long-haul link (LHL)} as a pair of consecutive IP addresses in a 
traceroute path separated by a latency that, in no-congestion scenarios, differs 
significantly from other latencies in the path. That is, in statistical terms 
the hop latency is an outlier.

We identify LHLs in large-scale traceroute campaigns and considered them independently 
of the underlying physical technologies connecting the consecutive hops (\ie physical mediums, 
link layer technologies). 

For our analysis, we focus on intercontinental LHLs, which we define as a LHL separated by
a set latency threshold (\S\ref{sec:method:threshold}) and where the pair of routers are in 
different continents. We focus on intercontinental connectivity assuming that this is the most 
complex part of the long-haul infrastructure and likely to be scarce partially due to deployment costs 
and its presence in multiple jurisdictions with potentially different legal and 
regulatory frameworks.

\subsection{Distances for long-haul links}
\label{sec:method:th}

In related work focusing on the domestic long-haul infrastructure of the US, Durairajan 
{\em et al.}~\cite{ram:intertubes} defines LHLs as those that connect major city-pairs, spanning 
at least 30 miles or connecting population centers of at least 100,000 people. We focus 
instead on intercontinental LHLs and our definition of LHL depends, instead, on a latency 
distribution of a measured path, a latency threshold and the location of both ends of LHLs. 

To identify a meaningful latency threshold for intercontinental LHLs we investigate distances between 
peering facilities as an estimator of link lengths. We focus on peering facilities as the importance of 
these infrastructures makes them strong candidates for end points of intercontinental LHLs.
Presence in these facilities enable direct peering with thousands of networks simultaneously.
While IXPs are typically meant to {\em keep local traffic local}, in recent years {\em remote peering}~\cite{remote:peering} 
has emerged as an approach to reach main content, transit and access networks.

\begin{figure}[th!] 
	\centering 
	\includegraphics[width=.38\textwidth]{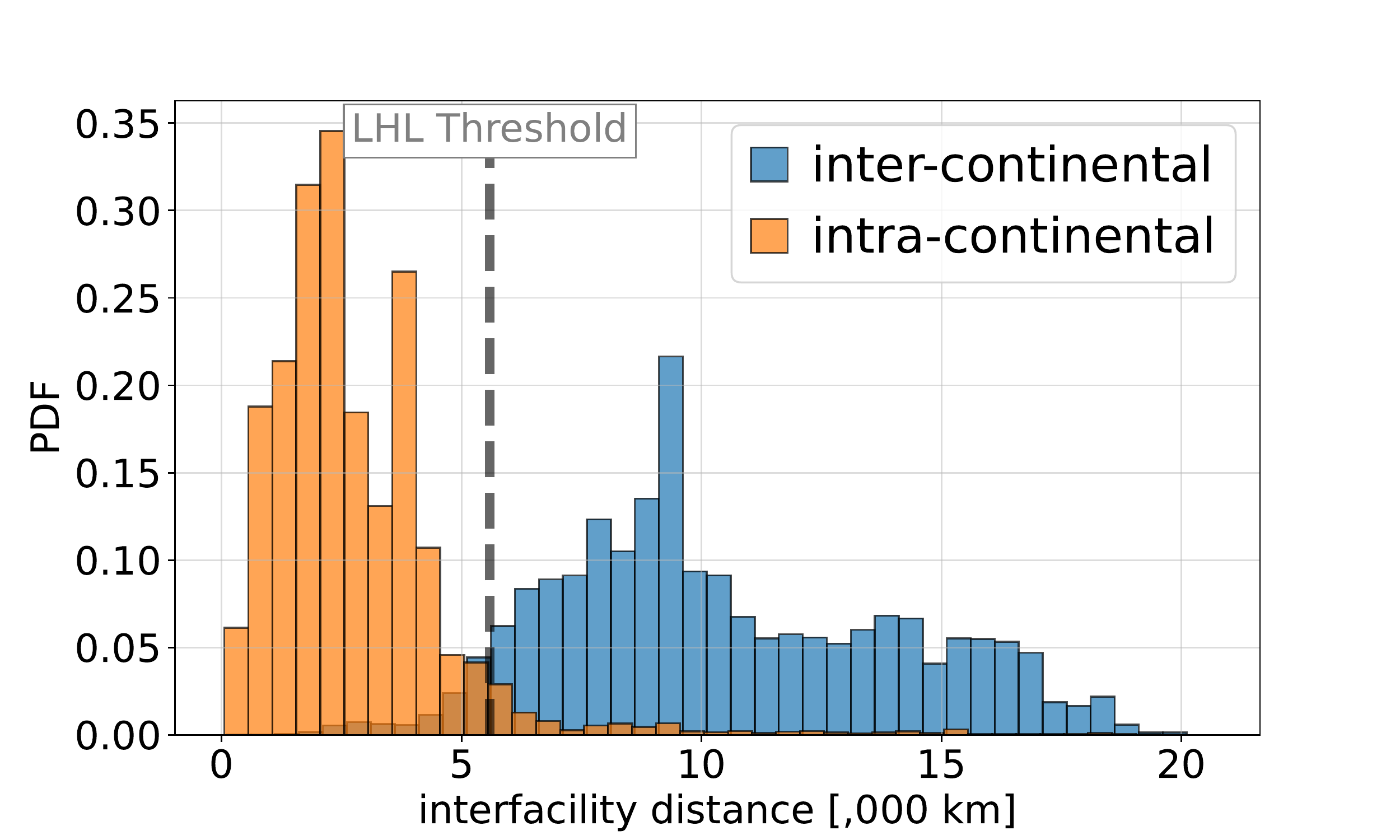}
	\caption{Histograms of intra-/inter-continental distances, in kilometers, 
		between the peering facilities listed on PeeringDB. 
		}
	\label{fig:cdf_interfacility_dist}
\end{figure}

We use a recent PeeringDB snapshot (October 2022) and investigate differences in inter- and intra-continental distance 
between all peering facilities to identify a suitable threshold for intercontinental LHL. We examine the distribution of 
distances between all pairs of networks present in different peering facilities (e.g., given two peering facilities with
two and four networks present at each, we have 8 distances). We use great-circle distance to compute the 
distance between each pair of peering facilities and repeat the process for each network pairs.
Figure~\ref{fig:cdf_interfacility_dist} shows histograms of the intra- and inter-continental distances (in kilometers) 
between {\em all} peering locations. We observe a shift in the intercontinental distribution with little overlap of both 
distributions after 5,700km. Nearly 95\% of networks are at a maximum distance of 5,657km from other networks present
at peering facilities in the same continent. Within that distance, there are only 5\% of networks reachable at 
inter-continental peering facilities.


Based on this analysis, in this analysis we set the LHL threshold of 5,700km or the equivalent of 57ms for RTTs 
propagating at $\frac{2}{3} \cdot c$ in traceroute measurements (assuming optical fibers and symmetric paths). 
This is a conservative lower bound since cable infrastructure is rarely deployed in straight lines~\cite{bozkurt2018dissecting}.
Thus, we restate our definition of intercontinental LHLs as \textit{a LHL with a latency of at least 57ms RTTs, 
where the pair of routers at each side of the LHL are in different continents.}

\section{Identifying Long-Haul Links}
\label{sec:method}

\begin{figure*}[th!] 
	\centering 
	\includegraphics[width=.9\textwidth]{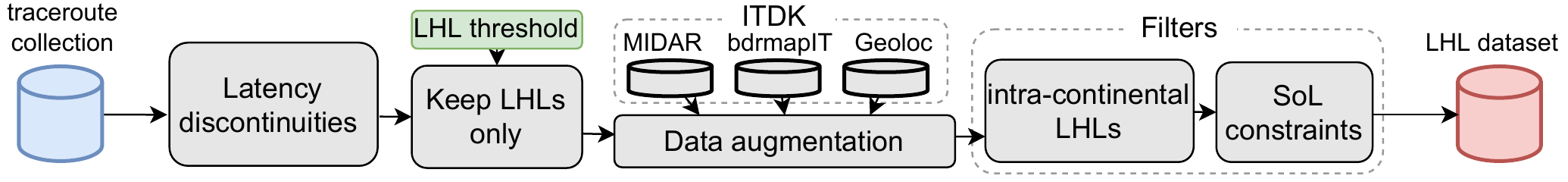}
	\caption{
	Identifying LHLs in traceroute measurements by $(1)$ detecting discontinuities in RTT 
	using anomaly detection, ($2$) adding topological information and selecting based on
	($3$) latency, ($4$) connecting intercontinental destinations and ($5$) with consistent 
	propagation delays.}
	\label{fig:pipeline}
\end{figure*}

To identify intercontinental LHLs in traceroute datasets, we first select 
candidate LHLs by detecting significant changes in RTTs (\S\ref{sec:anom_detection}), with 
latencies over {\em LHL threshold} (\S\ref{sec:method:threshold}). 
We then augment the list of candidate links with topological information 
(\S\ref{sec:method:data_aug}), and apply a two-stage filter to keep intercontinental 
LHLs using: router geolocation (\S\ref{sec:method:geoloc}), and a validation of inter-country 
distances with inter-hop latency values (\S\ref{sec:method:sol}). Figure~\ref{fig:pipeline} 
illustrates this process.

\subsection{Detecting discontinuities in traceroute}\label{sec:anom_detection}

We expect any given sequence of hops collected by traceroute to include at most one, 
and rarely more than one, LHL, making it easy to detect LHL candindates as large latency jumps.
A potential risk is the presence of false positives where significant lantecy jumps result 
from reasons other than propagation delay such as  queuing, latency 
inversions or the presence of middleboxes.

To identify LHL candiates as discontinuities in hop latency we frame the problem as anomaly 
detection and adopt commonly used tool for this task ADTK~\cite{adtk}. 
The ADTK \textit{LevelShift} inference method detects sequence shifts to a persistent state with higher values.
This method compares median values in two adjacent windows and identifies level shifts when the difference exceeds historical interquartile range adjust by a constant ($c=1$ in our implementation).
This feature 
allows us to exclude transitory latency changes while detecting increases of the propagation delay.
To gain further confidence in the links identified based on latency discontinuities, we conservatively keep 
links where both the previous and subsequent routers are responsive.

To illustrate the steps of our methodology we use the swath of the /24 IP space measured by one 
probe ({\tt jfk-us} in New York City, US) in one cycle (\textit{6537}, April 1, 2018) of the CAIDA 
Ark dataset. This dataset includes 56,637 traceroutes to addresses in different /24 networks. 
After applying our anomaly detection tool we are left with 
47,400
candidate LHLs out of 609,230 consecutive 
pairs of hops probed in
36,214 
traceroutes. Note that we use this slice of the dataset only 
for illustrative proposes and should not be seen as a representative sample of the full dataset 
(considering it is part of a single cycle captured by a single vantage point).
%

\subsection{Latency threshold}\label{sec:method:threshold}

The first step in the process of LHL identification yields a set of candidate
LHLs with latencies significantly different than their preceding and following hops. 

In this step we apply the {\em LHL threshold} (\S\ref{sec:method:th}) to candidate LHLs. 
We leverage multiple RTT samples, grouping all pairs of IP addresses connecting two routers. 
We identify the minimum RTT difference between router hops ($min_{diff}$), and then exclude 
all pairs of routers where the $min_{diff}$ is below the {\em LHLs threshold}.


Delay-based filtering has some clear limitations. Using the difference of RTT measurements between 
hops may result in inaccurate estimations of a hop latencies since packets expiring in different 
hops may have completely different reverse paths~\cite{steenbergen2009practical}. These estimations 
may also contain false positive inferences (\ie incorrectly include short-haul links) in the presence 
of latency inflation resulting from circuitous paths~\cite{katz2006towards}, diurnal congestion 
patterns~\cite{dhamdhere2018inferring} or misconfigurations. The next stages use additional 
information to address some of these limitations.

\subsection{Augmenting the dataset}\label{sec:method:data_aug}

Before selecting intercontinental LHLs from the set of candidates using different filtering 
conditions, we need to first augment the information associated with candidate LHLs using 
topological and geolocation information. 


We rely on CAIDA's Internet Topology Data Kit (ITDK)~\cite{itdk} for alias 
resolution~\cite{keys2012internet}, and to add router geolocation~\cite{maxmindgeoloc} and
router-to-AS mappings~\cite{marder2018pushing}. The ITDK is generated from a subset of the 
traceroute data gathered by Ark and it includes two related IPv4 router-level topologies, 
router-to-AS assignments, geographic location of each router, and reverse DNS lookups of all 
observed IP addresses~\cite{itdk}.

Using ITDK with the illustrative dataset, we can identify the ASN of 
6,553
routers 
(99.0\%) 
and geolocate 
3388 (98.46\%) 
of all routers.

\subsection{Removing intra-continental LHLs}\label{sec:method:geoloc}


We use router geolocation tags to select LHLs with end points in different continents. 
This step reduces the impact of latency jumps unrelated to propagation delay, including transient 
latency increments (\eg congestion and flash crowds) or other pathological events such as two
consecutive probes expiring in the same router, followed by a LHL. 

Router geolocation inferences are prone to contain errors~\cite{poese2011ip}, which we 
inherit from the ITDK dataset. We mitigate their impact by focusing most of our analyses 
at a country-level where geolocation services have been shown to provide better 
accuracy~\cite{huffaker2011geocompare, livadariu2020accuracy}.

After removing intra-continental LHLs from our dataset, we are left with 595 LHLs 
of the total set of candidate LHLs.

\subsection{Imposing speed-of-light constraints}\label{sec:method:sol}

The last step in our process of identifying intercontinental LHLs relies on 
Speed-of-Light (SoL) constraints, using minimum distances between countries, 
to address potentially incorrect geolocation inferences.

For this, we use countries' boundaries represented by polygons in the dataset 
to compute the nearest pair of points between {\em all} countries and obtain 
the minimum distance between {\em all} pairs of countries. We use these minimum 
distances to identify and remove long-haul links where inter-hop latency differences 
violate SoL constraints. 

After applying our complete process we are left with 571 intercontinental 
LHLs out of the 609,230 hops in our initial traceroute dataset.

%% file: tex/dataset.tex
\section{Dataset}\label{sec:dataset}

Our methodology for LHL identification can be applied to any large-scale traceroute dataset. 
In this section, we describe the specific datasets we use for our analysis of intercontinental long haul connectivity. 

\begin{table}
\centering

\caption{Snapshots of traceroute measurement campaigns collected by CAIDA's Ark platform.}\label{table:dataset}
\footnotesize
\begin{tabular}{cccc}
\toprule
year		& cycles	& \# probes & \# traceroutes\\
\hline
\hline
2016		& 4576, 4577, 4578		& 97	&32.72M\\	
2017		& 5422, 5423, 5424		&117	& 33.08M\\
2018		& 6446, 6447, 6448		&149	& 32.71M\\
2019		& 7615, 7616, 7617		&112	& 32.72M\\
2020		& 8820, 8821, 8822		&121	& 33.99M\\
2021		& 9643, 9644, 9645		&69	& 32.84M\\
2022		& 10019, 10020, 10021	& 95	& 32.83M\\
\hline
7 years	& 21 					&244	&231.45M	\\
\bottomrule
\end{tabular}

\end{table}

\subsection{Traceroute measurements}

The core of our analyses leverages traceroute measurement campaigns collected by the CAIDA's Archipelago 
(Ark) platform~\cite{ark}. Ark's measurement campaigns are Internet-wide topological explorations that 
use a /24 granularity to cover all IPv4 prefixes announced in BGP routing tables~\cite{beverly2010primitives}. 
In each traceroute campaign, or {\em cycle}, all /24 subnets are probed from one vantage 
point~\cite{caida-probing-ipv4}.

Table~\ref{table:dataset} shows the measurement cycles included in our analysis with details on the number of 
vantage points and traceroute measurements. We combine three consecutive measurement cycles of the same day 
(\eg 6446-6448 for 2018) to capture links that may have been missed due to packet loss and to help identify and 
remove transitory latency  inflation with additional RTT samples. 

We use the most recent cycle, cycle 2022 (1) (10019-10020-10021), to study the long-haul infrastructure today 
(\S\ref{sec:observations}). For the longitudinal part of our analysis (\S\ref{sec:longitudinal}), we use data 
collected over a period of 7 years starting in 2016. 

As described in (\S\ref{sec:method}), we use ADTK LevelShift~\cite{adtk} to identify discontinuities in these 
datasets. LevelShift is compute intensive and  we manage its CPU impact at scale by randomly picking 1/10 of the 
traceroute measurements in a given dataset. This random down-sampling should not impact the visibility of LHLs 
in the Internet core which should be present in multiple measurements~\cite{Xun2010Selecting, caida-pam-2001}.

\subsection{Complementary datasets}

As noted in Section~\ref{sec:method}, we augment our long-haul inferences with additional topological information 
from CAIDA's ITDK kit~\cite{itdk}.  
To examine the data at different granularities, we include 
three topological features to our inferences: \ione IP-to-router mappings using MaxMind-based 
geolocation inferences~\cite{maxmindgeoloc}, \itwo router-to-country geolocation inferences using 
bdrmapIT~\cite{marder2018pushing}, and \ithree router-to-AS mappings using MIDAR alias 
resolution~\cite{keys2012internet}. We also include PTR records for the IP addresses (router hostnames) 
in the long-haul link data collection since 
this information can provide insights on geolocation~\cite{rnds-geoloc},
customers~\cite{rdns-learning} or other relevant 
features about the network deployments and structure.

We use daily PeeringDB snapshots, collected during the days of the measurement cycles, to investigate 
characteristics of peering ecosystems as a driver of intercontinental long-haul connectivity. 
Despite inaccuracies in PeeringDB records, and its limited and potentially biased coverage, previous 
studies have shown that it includes a representative picture of the network~\cite{lodhi2014using, kloti2016comparative}.
Last, we also rely on other datasets for our analysis including CAIDA's AS relationship files~\cite{asrel, giotsas2014inferring}, 
CAIDA's IXP dataset~\cite{caida-ixp}, and geographic information~\cite{mayer2011notes}.

%% file: tex/observations.tex
\section{Long-haul connectivity today}
\label{sec:observations}

We begin our study of the intercontinental long-haul infrastructure using the most 
recent data from Ark, with cycle 2022 (1) (10019-10020-10021). We explore the 
characteristics of LHLs, their preferred destinations, and the resulting graph. 

\subsection{LHLs: Lengths and destinations}
\label{sec:observations:ulh}

\begin{figure}[th!] 
	\centering 
	\includegraphics[width=.22\textwidth]{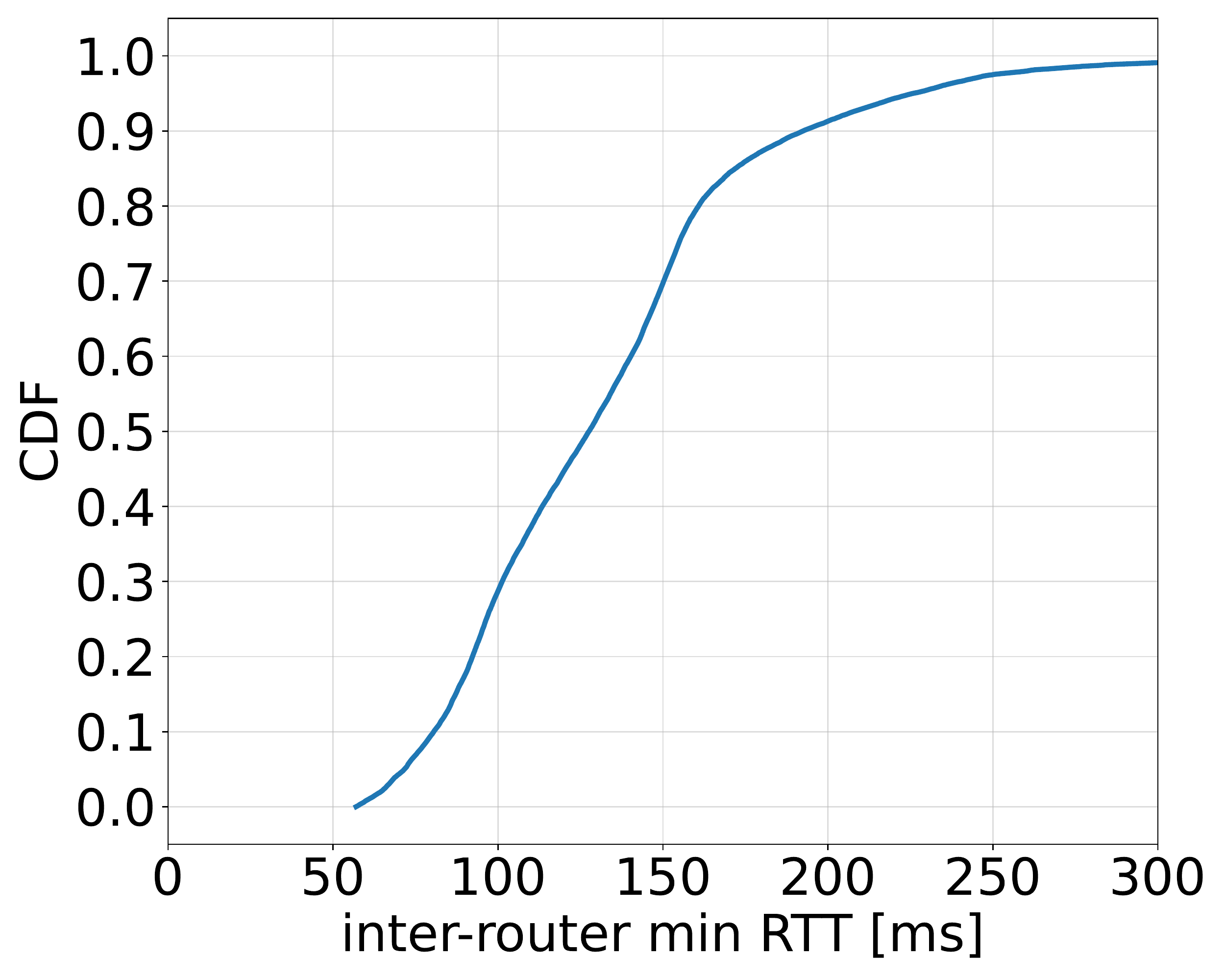}
	\caption{
		Cumulative distribution of long-haul inter-router latency. 
		}
	\label{fig:lhl_rtts:all}
\end{figure}

The LHL identification process (\S\ref{sec:method}) finds 31,452 LHLs connecting 
15,031 routers in 146 countries. To characterize LHLs, we begin by considering the 
distribution of LHL latencies. Figure~\ref{fig:lhl_rtts:all} is a CDF of inter-router 
latency differences (in milliseconds) and shows a somewhat steep distribution with 75\% 
of LHLs having latencies between $\sim$60ms and $\sim$155ms. The distribution flattens at 160ms in long tail for the last 20\% of the LHLs.




\begin{figure}[th!] 
	\centering 
	\includegraphics[width=.38\textwidth]{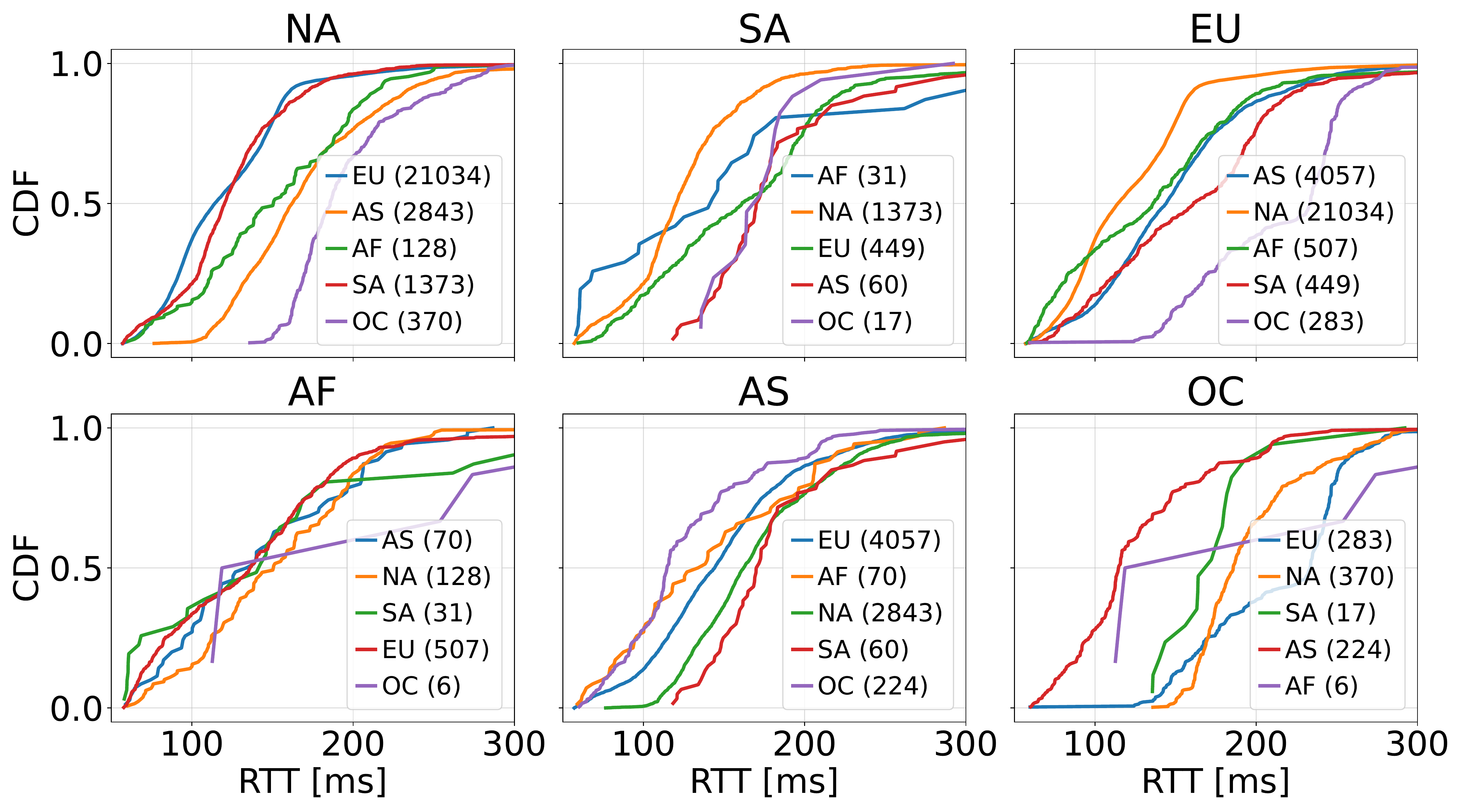}
	\caption{Inter-router latency distribution.}
	\label{fig:lhl_rtts:conts}
\end{figure}

To provide a perspective of individual inter-router latency differences between continent pairs, we plot a per-continent breakdown in Figure~\ref{fig:lhl_rtts:conts}.
This figure shows us that, to a different extents (prevalence is shown in brackets), all continents have LHLs connecting each other.
We observe that North America connects to Asia and Oceania where speed-of-light constraints gaps are visible in the distribution, likewise in South America to Asia and Oceania and Europe to Oceania.

The reach of some of the identified LHLs is quite impressive, including single hops connecting distant locations such as 
Los Angeles (US) and Budapest (HU) or Sao Paulo (BR) and Tokyo (JP), some of them up to $\sim$20,000km apart.
This is a clear sign of decoupling between the physical infrastructure and the network layer. There is no submarine 
infrastructure directly connecting some of these points, such as South America and Asia, but instead the concatenation as
single network-level hops of multiple physical segments through virtualization mechanisms.



\subsection{Visible MPLS in LHLs}

While it is not possible to characterize the adoption of all virtualization mechanisms, 
in the following paragraph we investigate the fraction of LHLs contain visible MPLS tunnels 
from traceroute data. 

The visibility of MPLS tunnels depends on the combination of RFC4950-compliant nodes and ingress 
nodes enabling the {\tt ttl-propagate} option~\cite{donnet:mpls}. RFC4950-compliant nodes 
append the MPLS label stack of the time-exceeded message to the ICMP packet providing traceroute 
visibility to Label Switching Routers (LSR) of the Label Switched Path (LSP). If the first MPLS router 
of an LSP copies the IP-TTL value to the LSE-TTL field rather than setting the LSE-TTL to an 
arbitrary value such as 255, LSRs along the LSP will reveal themselves via ICMP messages even 
if they do not implement RFC4950. If MPLS nodes implement both RFC4950 and {\tt ttl-propagation}, 
MPLS tunnels are called {\em explicit tunnels} while if they implement RFC4950 but not {\tt ttl-propagation}, they
are referred to as {\em opaque tunnels} where only the last hop of the LSP is visible.

We can identify the presence of MPLS tunnels in LHLs when the ICMP time-exceeded message of far-side 
node contains MPLS labels in the payload. 
More sophisticated inferring techniques can improve visibility of {\em invisible tunnels}~\cite{vanaubel2017through}.
In this work we investigate the most elemental implementation of MPLS tunnels leaving a more exhaustive explorations for future work.


\begin{figure}[th!] \centering 
	\includegraphics[width=.38\textwidth]{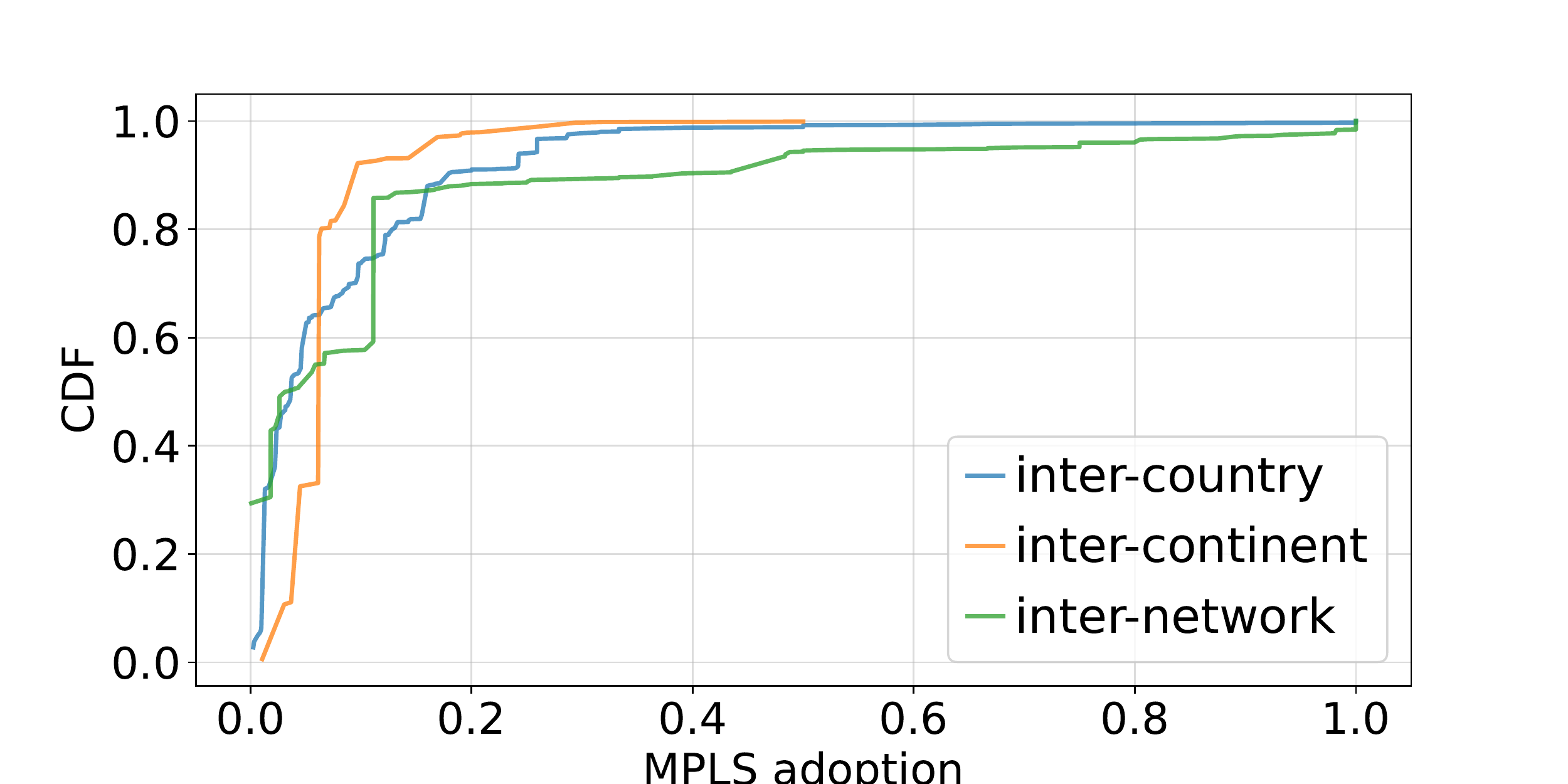}
	\caption{
		Weighted CDF of the MPLS adoption at the country, continent and AS levels. 
		The weights are given by the number of LHLs between pairs.
	} \label{fig:mpls}
\end{figure}

We find evidence of the use of MPLS in 1,932 LHLs ($\approx$ 6.14\% of all LHLs)  to connect 2,208 routers ($\approx$7.02\%).
We look at the prevalence of  LHLs over visible MPLS tunnels from country-, continent- and AS-level perspective. 
Figure~\ref{fig:mpls} shows the weighted CDF MPLS adoption in LHLs at the country, continent and AS levels where the weights are given by the number of LHLs between pairs.
These curves show that the prevalence of this type of tunnels is low with weighted average values of 7.79\%, 6.86\%, 11.89\% at the country, continent and AS levels, respectively.
At the AS level, the prevalence of these tunnels is far from being homogeneous across networks finding that a remarkable adoption in some specific networks, such Claro Brazil (AS4230), NTT (AS2914), Uruguay's ANTEL (AS6057), or as Vodafone (AS1273), where the adoption ranges, respectively, between 75.0\% (108/144), 80.5\% (70/87), 81.25\% (13/16),  and 98.2\% (111/113).


\subsection{A compass view of LHLs}\label{sec:compass}

As a last step in our analysis of LHLs, we investigate whether there is a preferred orientation in the 
deployment and use of intercontinental long-haul infrastructure. We assume that, if only for historical reasons, the majority 
of the underlying infrastructure behind LHLs will be oriented East-West. That said, the rapid growth of Internet 
connectivity in the southern hemisphere (\eg Brazil, Oceania) may mean a growing number of LHLs supporting 
connections to the large infrastructure hosted predominately in the northern hemisphere.

\begin{figure}[th!] 
	\centering 
	\includegraphics[width=.35\textwidth]{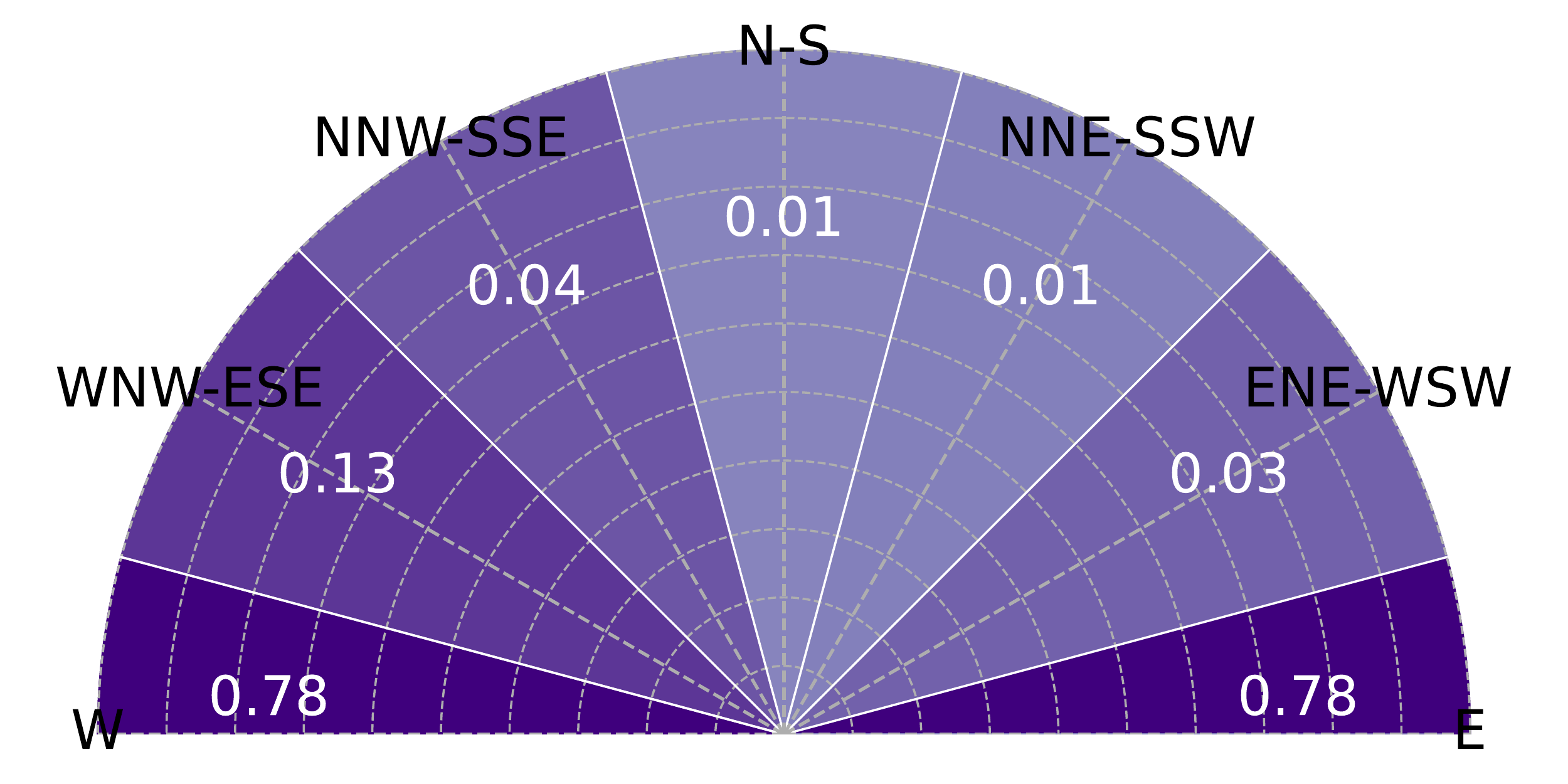}
	\caption{
		Prevalence of inter-router connections directions in the 12-wind compass rose.
	}
	\label{fig:compass}
\end{figure}

Figure~\ref{fig:compass} shows a semicircle with 6 parts representing the 6 directions in which two pairs of coordinates can be in a 12-wind rose 
(\eg North-South (N-S), North-Northeast to South Southwest (NNE-SSW)). In each direction, we list the fraction of LHLs in this dataset (note that E-W includes the same number in each side of the rose). As we suspected, we observe that the vast majority of the long-haul links (78\%) connects routers in the East-West direction, where the most prevalent inter-country links in this direction correspond to North Atlantic connectivity between the US and European counterparts including Germany, France, the UK and the Netherlands (\S\ref{sec:prefferred_destination}). We suspect that the large availability of submarine infrastructure in the North Atlantic~\cite{jyothi2021solar} with a highly diversified and competitive market that includes dozens of operators (\eg Zayo, TATA, Orange, Level3, Hurricane Electric, Deutsche Telekom, \etc) is behind the large fraction of connectivity in this direction. The North-South connectivity, that partially includes the connectivity between Northern and Southern hemispheres, at $\approx$1\% clearly lags by comparison with other orientations. While this orientation includes connectivity between the US and Argentina and Chile, and between the Netherlands and the UK with South Africa, the southern countries also have many intercontinental connections in NNE-SSW and NW-SE directions.

\subsection{Preferred long-haul destinations}
\label{sec:prefferred_destination}

We now change focus to the target of LHLs and investigate the preferred destinations of long-haul connections and whether preferences vary across continents. Different factors are likely to shape long-haul preferences. These include technical issues such as content availability and peering opportunities, geographic features challenging long-haul deployments (\eg being able to anchor landing points, presence of strong water currents), cultural affinity (\eg a common language) and economic cooperation (\eg the European Union partially financed the EASSy cable in East Africa~\cite{eassy}).

\begin{figure}[th!] 
	\centering 
	\includegraphics[width=.4\textwidth]{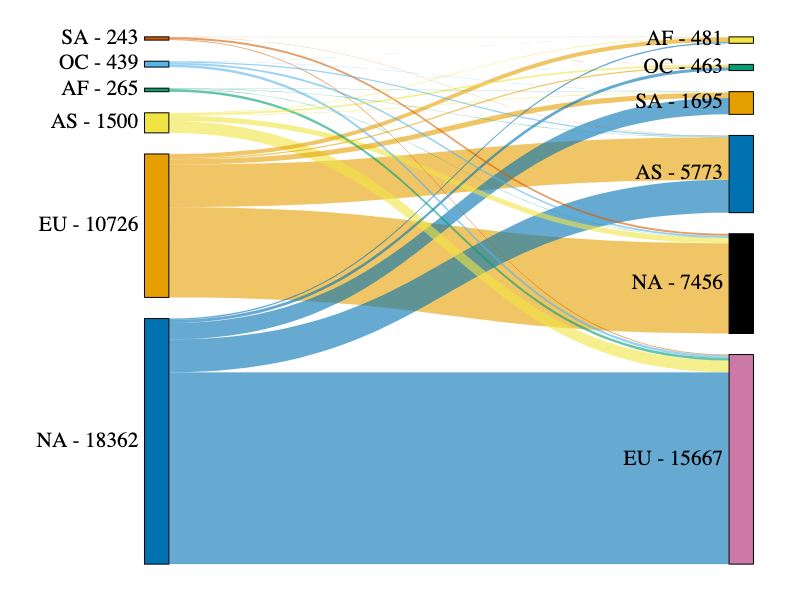}
	\caption{Intercontinental long-haul connectivity.}
	\label{fig:matrix}
\end{figure}

We start by looking at preferred long-haul destinations at continent-level granularity.
Figure~\ref{fig:matrix} show a Sankey diagram of the prevalence of  LHLs between continents.
The large majority of LHLs have the near-side router in North America, with far-side routers in all regions, but most commonly in Europe.
From the far side perspective, North America is the major contributor to LHLs terminating in Europa and South America and a remarkable actor in the rest of the regions.
We find a correlation between these preferences and the number of submarine cable connecting continents. For instance the number of submarine cables that connects Asia with Europe (28) doubles the number that connects Asia with North America (North America).
The North Atlantic routes dominates intercontinental LHL connectivity with $\approx$67\% (21033) of all LHLs in this snapshot.

\begin{figure}[th!] \centering 
	\subcaptionbox{\label{fig:preferred_foreign:1}}{%
		\includegraphics[width=.14\textwidth]{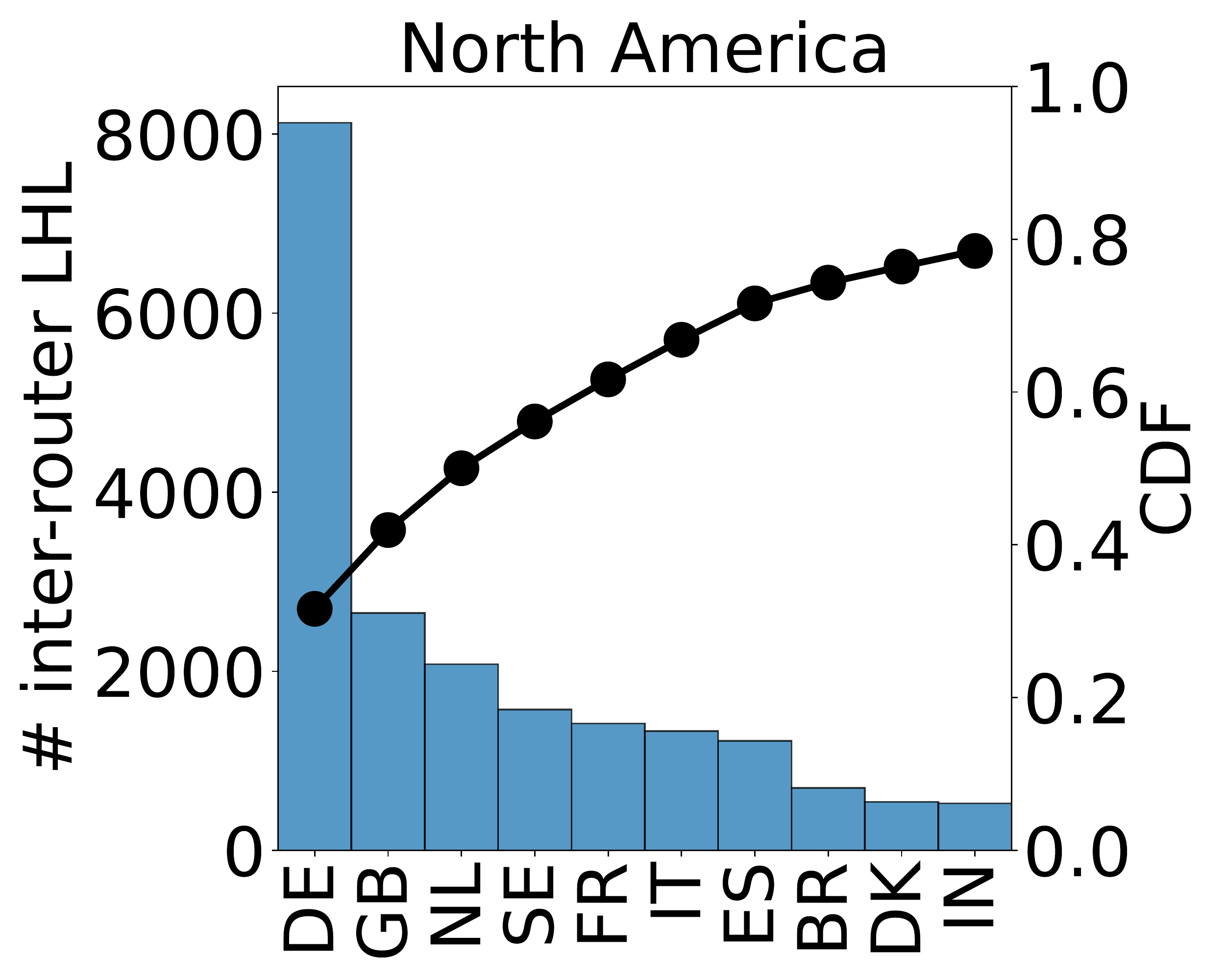}
	}~\subcaptionbox{\label{fig:preferred_foreign:2}}{%
		\includegraphics[width=.14\textwidth]{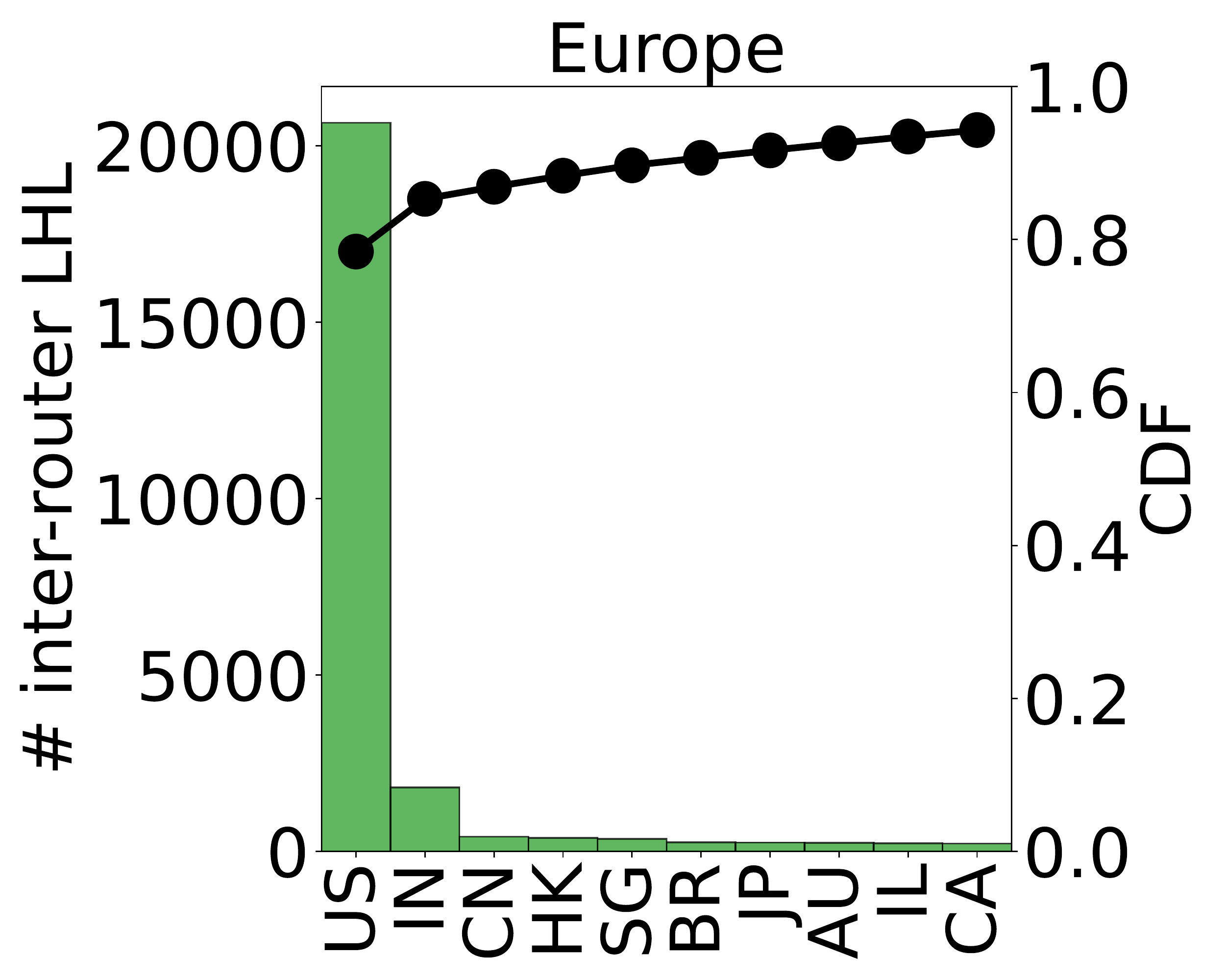}
	}~\subcaptionbox{\label{fig:preferred_foreign:3}}{%
		\includegraphics[width=.14\textwidth]{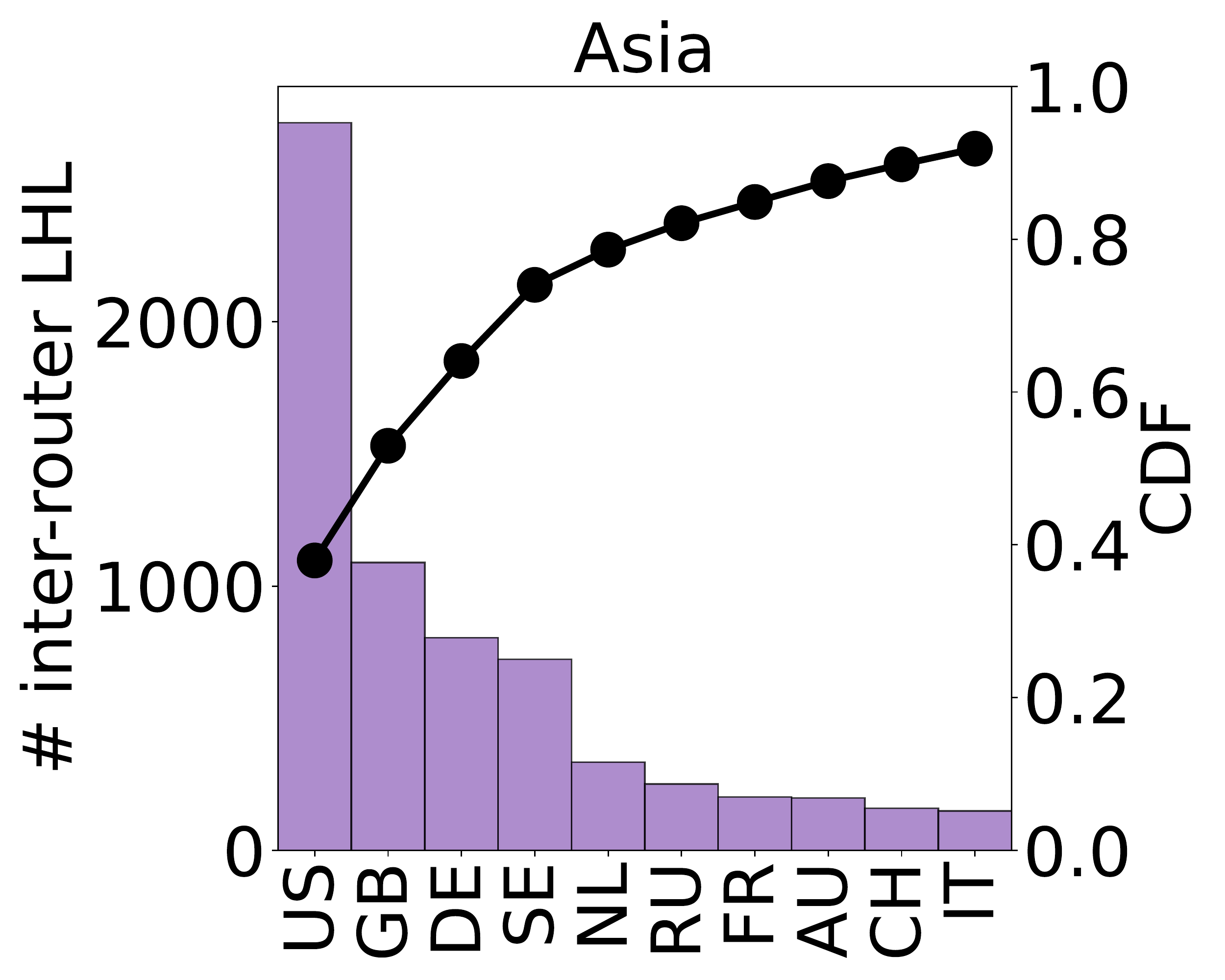}
	}\\
	\subcaptionbox{\label{fig:preferred_foreign:4}}{%
		\includegraphics[width=.14\textwidth]{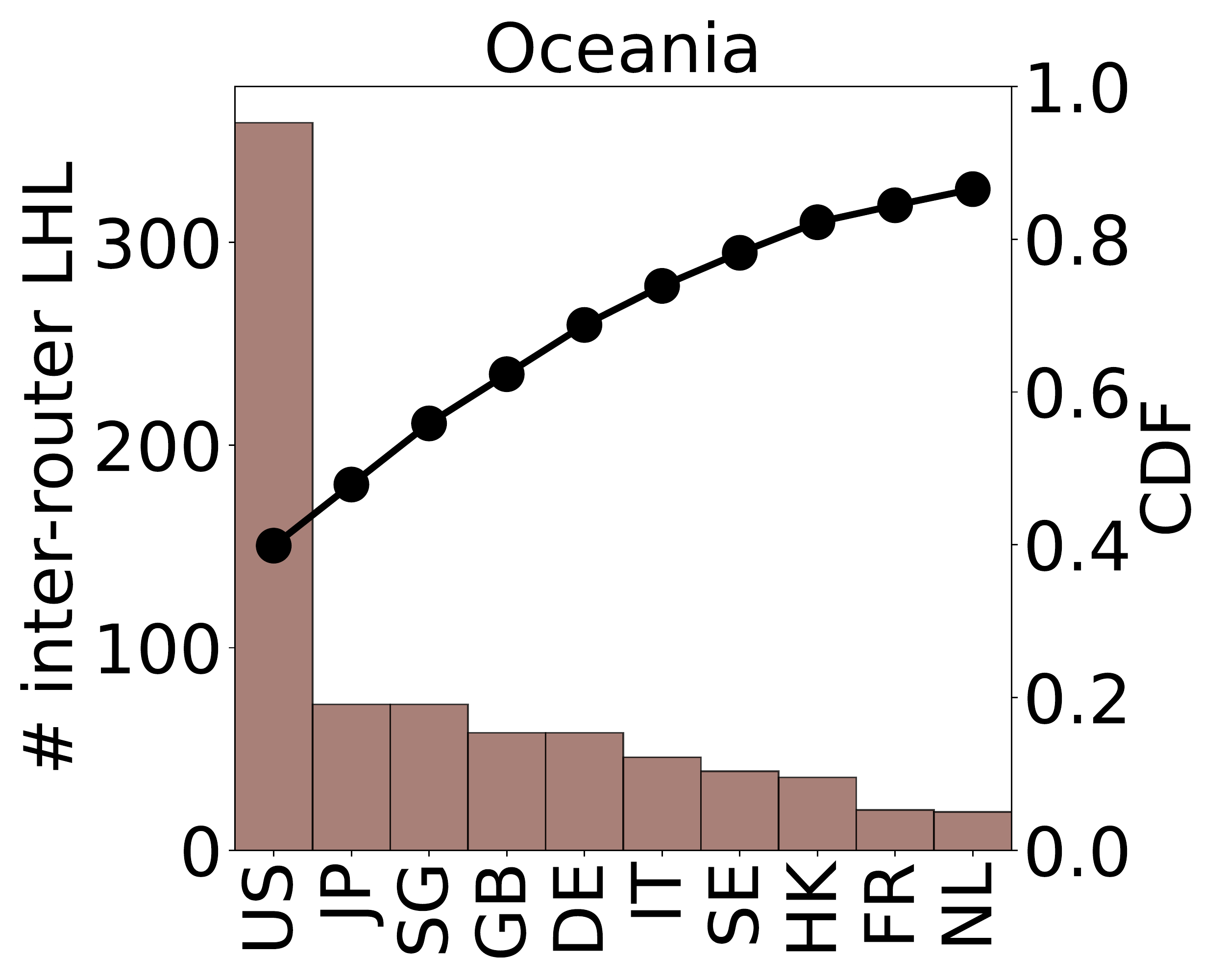}
	}~\subcaptionbox{\label{fig:preferred_foreign:5}}{%
		\includegraphics[width=.14\textwidth]{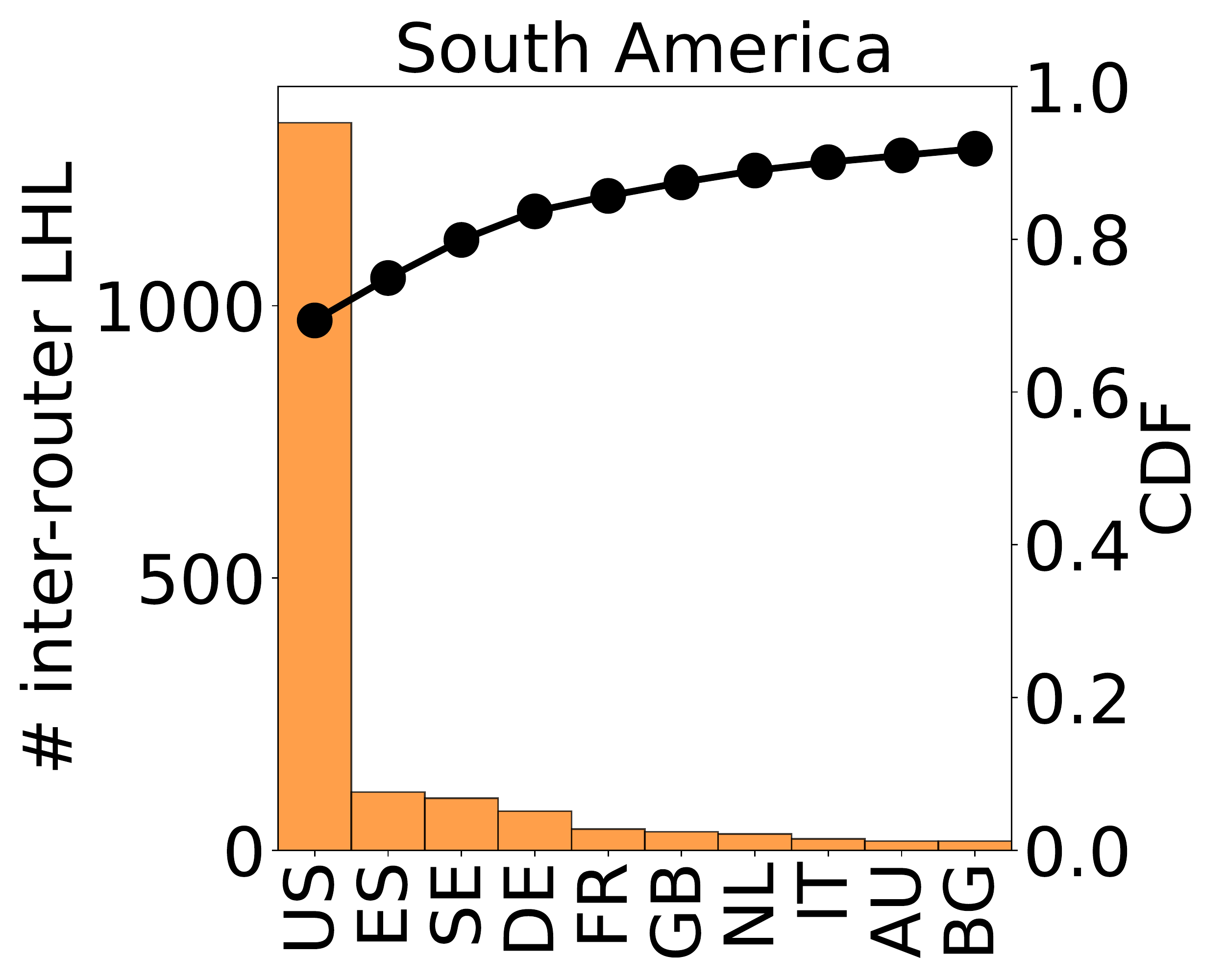}
	}~\subcaptionbox{\label{fig:preferred_foreign:6}}{%
		\includegraphics[width=.14\textwidth]{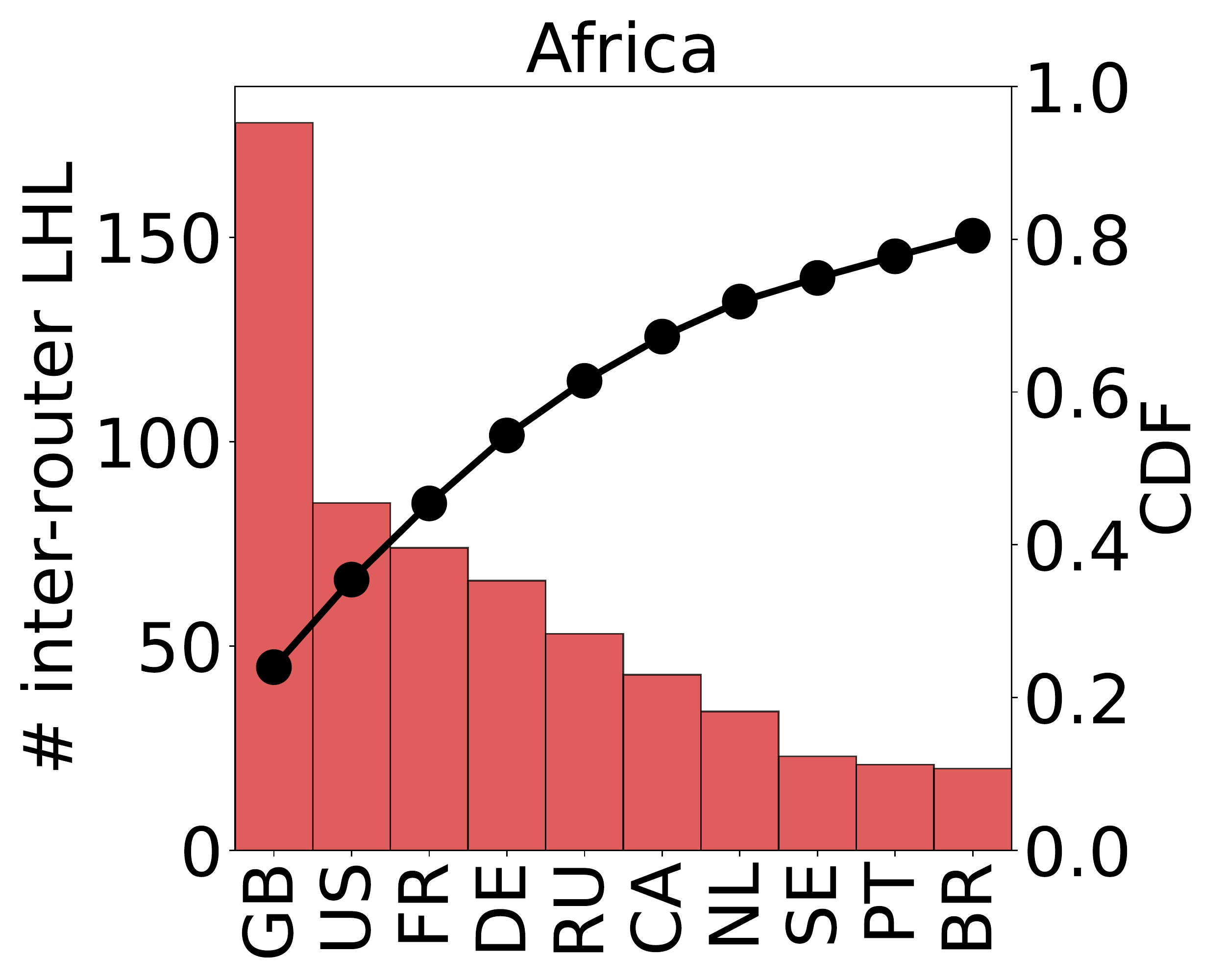}
	}
	\caption{
	Preferred destinations for long-haul connectivity across regions. 
	} \label{fig:preferred_foreign}
\end{figure}

We now shift our attention to preferred long-haul destination at a country level.
Figure~\ref{fig:preferred_foreign} shows the top 10 preferred countries for long-haul connectivity across different regions. The vertical axis of each figure shows the number of LHLs while the horizontal axis lists the top 10 countries, in order, among the preferred destination for a given region. 
We observe the US as the preferred destination for all regions, except North America and Africa, where the fraction of {\em all} inter-router long-haul links ending in the US drops to 0.23.

The preference of African countries for major European hubs over the US may be motivated by proximity and other factors such as cultural affinity. Previous studies have shown, for instance, the prevalence of French operators across French-speaking countries in Africa~\cite{fanou2015diversity} and the presence of circuitous paths in African connectivity including detours to London and Amsterdam~\cite{gupta2014peering}. A similar cultural influence is visible in South America where the second preferred destination is Spain. Another observation is that despite efforts, a large number of countries rely on content that is hosted in or routed through North America~\cite{edmundson2018nation, jacquemart2019inferring}. Other countries common to all regions top-10 listing includes Singapore, Great Britain, Germany and France. Singapore is a major hub with a large, state-owned transit ISP SingTel (AS7473)~\cite{carisimo2021identifying} while Great Britain, Germany and France host some of the largest IXPs in the world (LINX, DE-CIX and France-IXP).

\subsection{Super routers}
\label{sec:superrouters}

We look at the end points of the LHLs, their node degree, and geographic footprint. 

\begin{figure}[ht!] \centering 
	\subcaptionbox{\label{fig:node_degree}}{%
		\includegraphics[width=.16\textwidth]{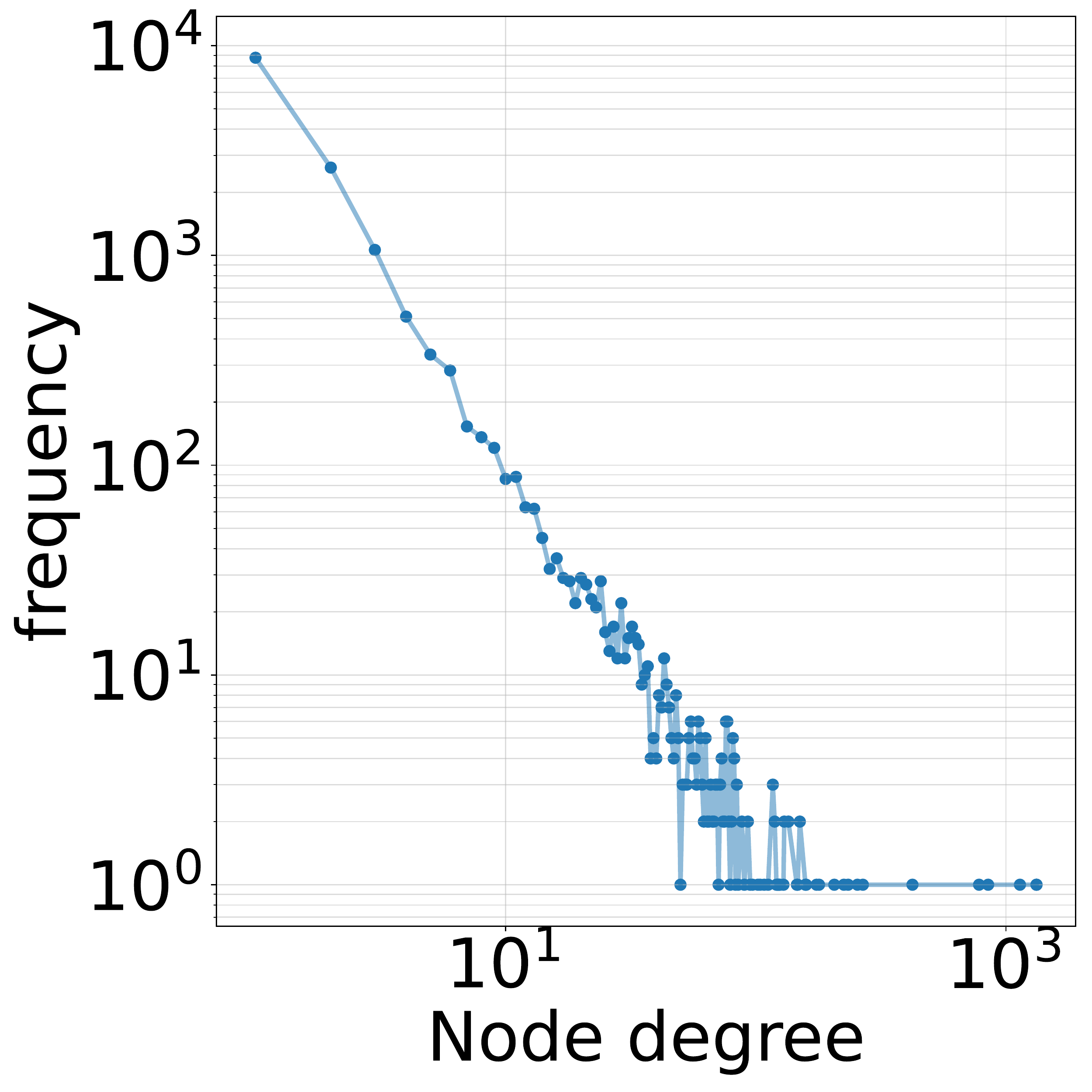}
	}~\subcaptionbox{\label{fig:cc_cdf:1}}{%
		\includegraphics[width=.16\textwidth]{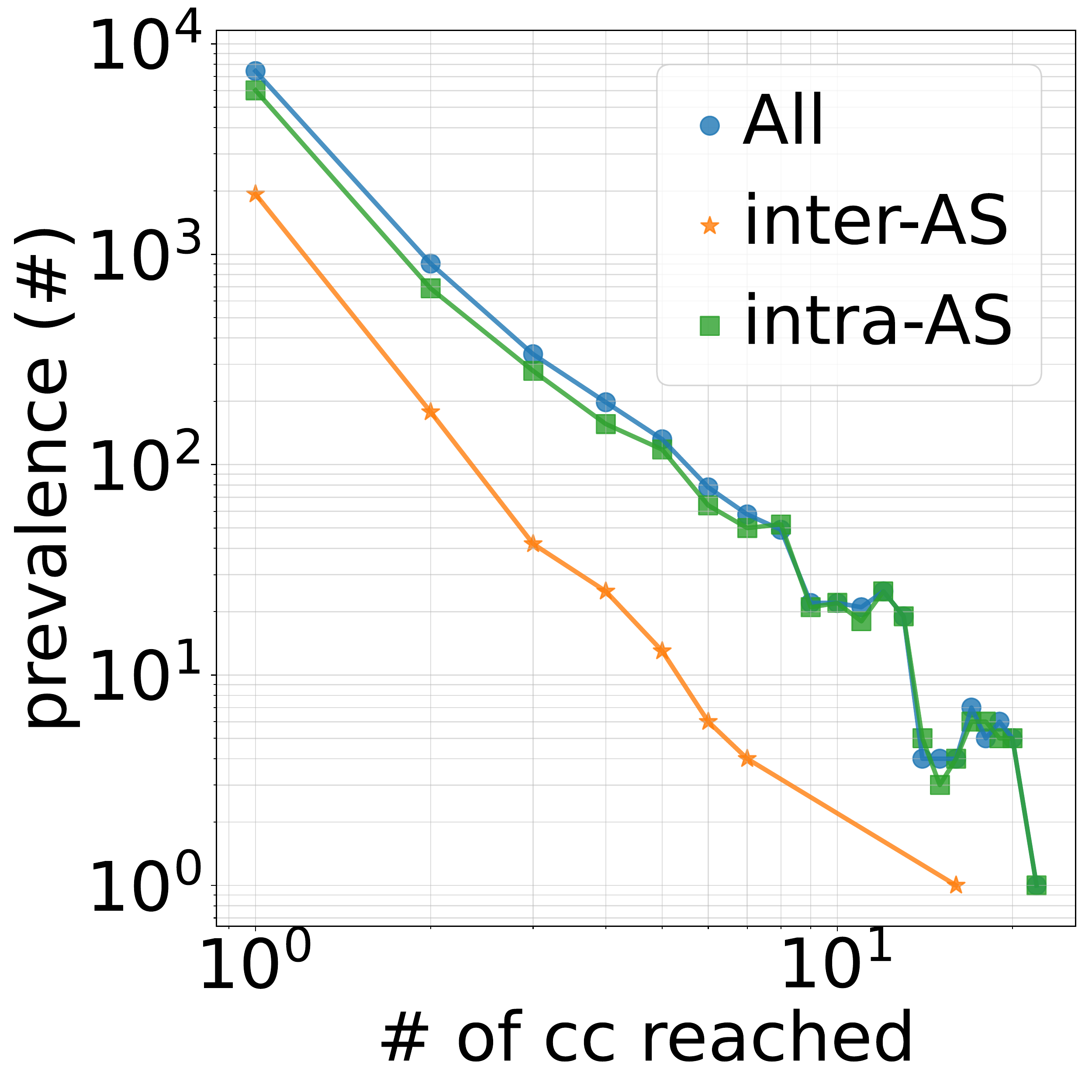}
	}~\subcaptionbox{\label{fig:cc_cdf:2}}{%
		\includegraphics[width=.15\textwidth]{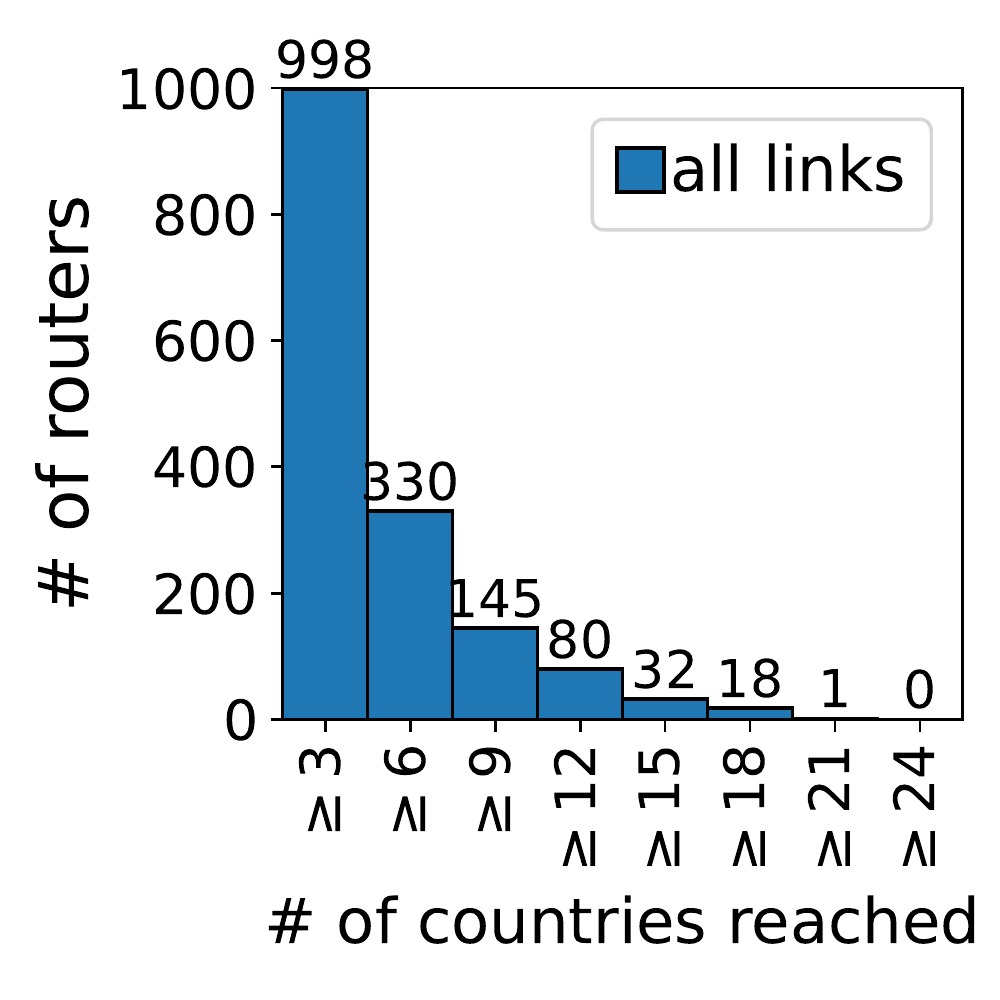}
	}
	\caption{
	Node degree distribution (Fig.~\ref{fig:node_degree}) and 
	country-level footprint of routers shown as a CCDF (Fig.~\ref{fig:cc_cdf:1}) and bar graph (Fig.~\ref{fig:cc_cdf:2}) .
	} \label{fig:cc_cdf}
\end{figure}

Figure~\ref{fig:node_degree} shows the node degree distribution of these end points. 
As the figure shows, node degrees range widely from a handful to as many as 1,326, and 
a clear long-tail distribution with the top 5$^{th}$ percentile of vertices having node degrees 
larger than 13. Changing perspectives to the country-level connectivity of vertices, we find 
the top 5$^{th}$ percentile connecting between 1 and 24 countries!

\begin{figure}[ht!] \centering 
	\includegraphics[width=.38\textwidth]{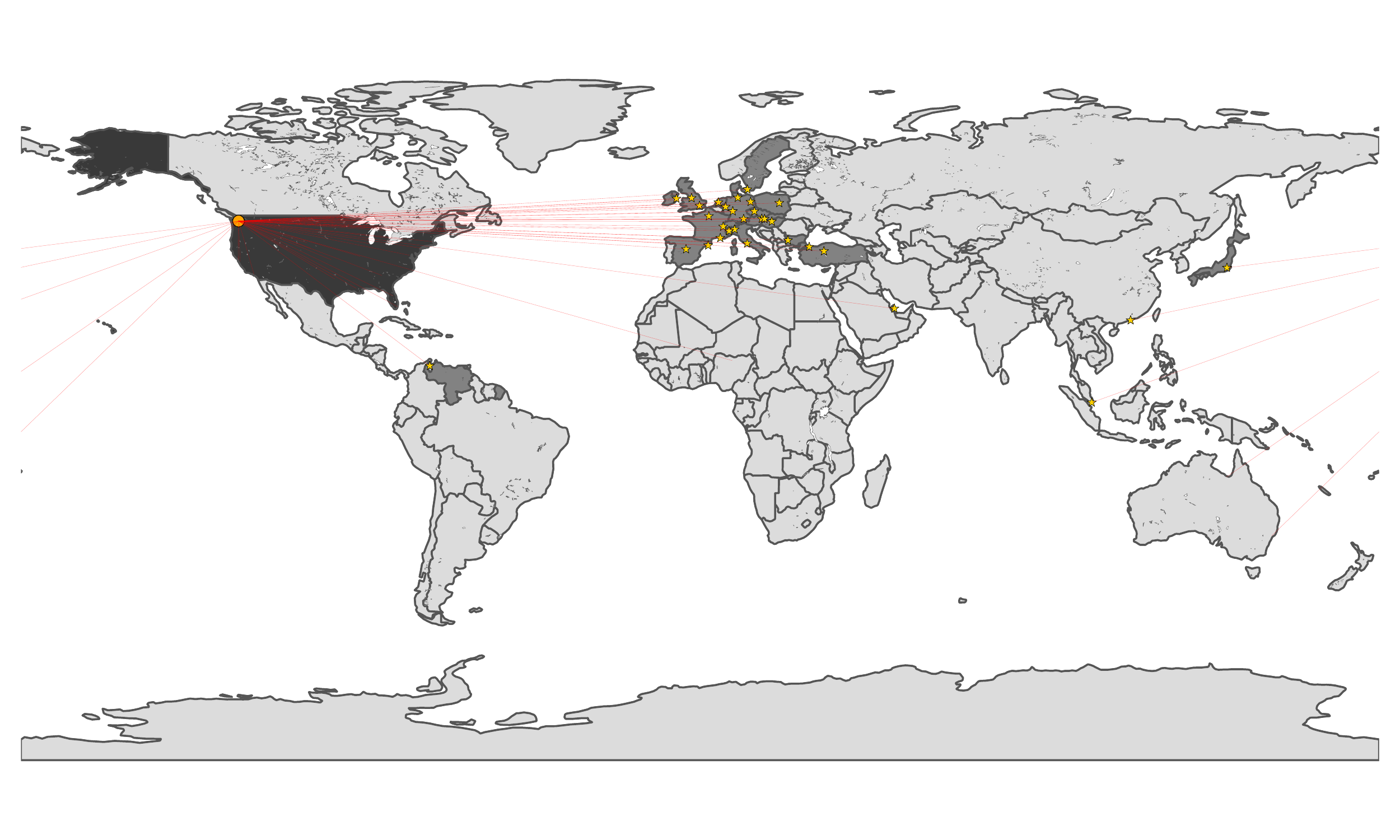}
	\caption{An example of a super router (GTT, AS3257).}
	\label{fig:superrouters}
\end{figure}

We call this popular LHLs destinations \textit{super routers}, given their large router and country-level 
reach. Specifically, we define a {\em super router}\footnote{Appendix~\ref{sec:appendix:tr:superrouters} 
shows raw traceroute sequences traversing a {\em super router}.} as those routers with
directly connected (\ie next hop) with large number of routers scattered across several countries. 
Figure~\ref{fig:superrouters} illustrates the idea of {\em super routers} with an example of a router 
operated by GTT (AS3527) in Seattle, Washington connecting neighboring routers scattered in 23 countries.

We investigate the geographic distribution of {\em super routers}, focusing on their location
and country-level footprint. 

Given the preferred orientation of LHLs (East-West), some of the popular destinations at 
either end of LHLs, as well as the distribution of their length, one would imagine LHLs 
to be the \textit{network-level view} of submarine cables links and thus to have end-points, primarily, 
in coastal areas, more or less evenly distributed near submarine cables' landing points. While 
there are instances of LHLs matching this intuition, our analysis shows that many popular routers 
are found in-land -- as far in-land as Chicago (US), over 700km (as the crow flies) from the 
closest landing point (Tuckerton, US)!  

The country-level footprint of routers associated with LHLs is shown in Fig.~\ref{fig:cc_cdf:1}. The graph 
plots the CCDF of the number of countries that each router is connecting to via LHLs. While 90\% of super-routers 
connect at most  3 countries, the top 1\%, 2\% and 5\% super routers connect at least to 11, 8 and 4 countries, 
respectively. Among ASes, we note the vast majority operating routers connecting to 3 or 6 
countries, with a selective group of 11 ASes (mainly Tier-1 transit providers) operating super routers 
as international gateways connecting 9 or more countries.

\begin{figure}[th!] 
	\centering 
	\includegraphics[width=.38\textwidth]{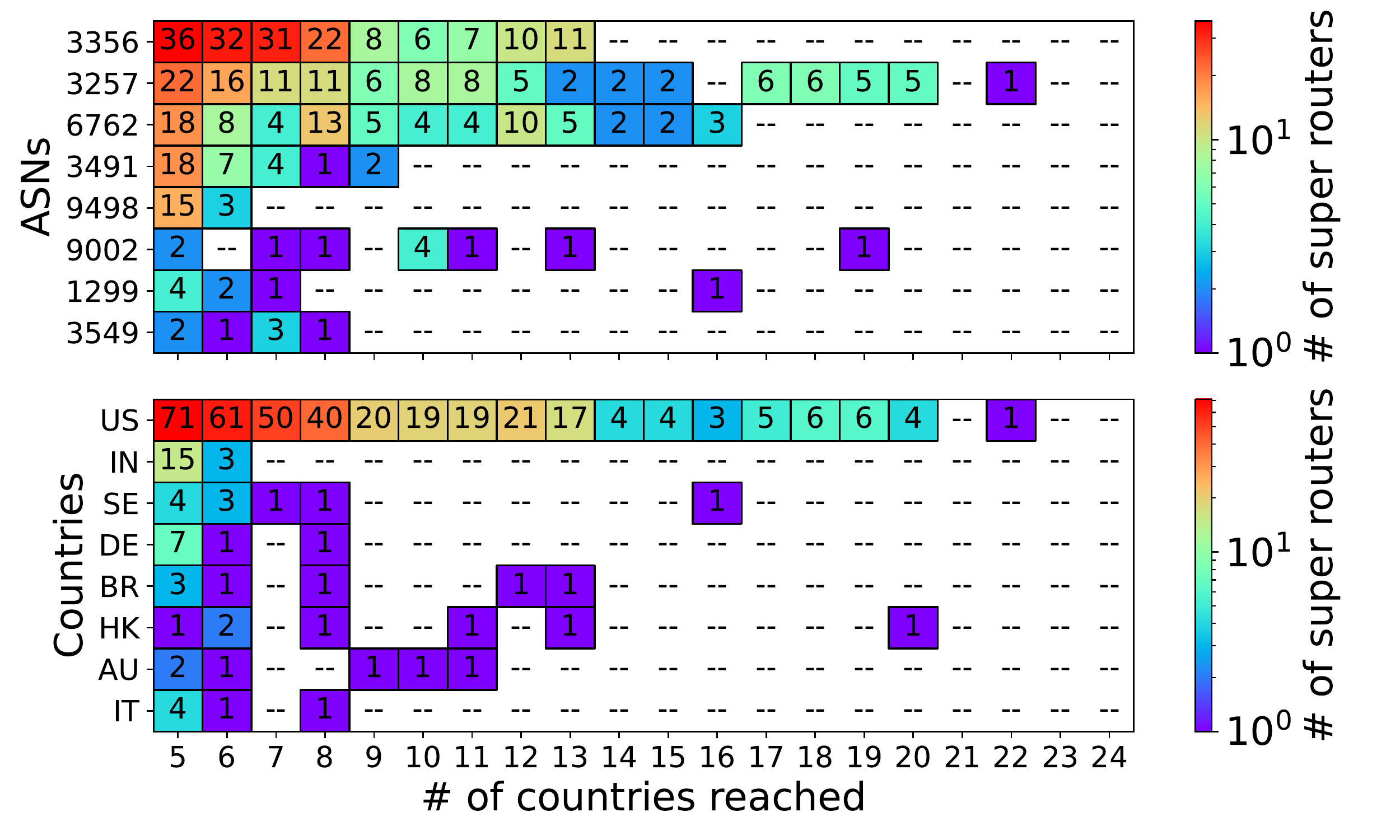}
	\caption{Top 8 ASes and countries with {\em super routers} connecting at least 5 countries}
	\label{fig:countries_ases_leaders_in_use_of_superrouters}
\end{figure}

Figure~\ref{fig:countries_ases_leaders_in_use_of_superrouters} shows the top 8 ASes (top) and countries 
(bottom) holding the largest number of super-routers connecting 5 countries or more via LHLs. We observe 
that international Tier-1 transit providers lead in the use of {super routers} with AS3356 (Lumen, formerly Level3) 
AS3257 (GTT), AS6762 (Telecom Italia) and AS3491 (PCCW) at the top of this list.

From a country-level perspective, the US hosts by far the largest number of super router,
followed by India, Sweden and Germany. The presence of super routers in Germany can be explained, at least in part, 
by the scale of the IXPs it hosts,  Other popular countries are headquarters of large transit networks 
(Telecom Italia-AS6762 and Arelion-AS1299 in Italy and Sweden, respectively).  The dominance of the US is perhaps not 
surprising given that the country hosts over a quarter of locations for many of the top cloud computing services such as 
Google Cloud Platform, Microsoft's Azure and IBM (25.7\%, 25\%, 27.3\%, respectively), and over half of Amazon's 
locations (52.6\%), and is also relatively far from the rest of the world.

\subsection{Takeaways}

In this section, we presented key observations of the intercontinental long-haul connectivity.
We explored the inter-router delay, inter-country distance and orientation of LHLs. We 
found that a quarter of LHLs have at least 155ms delay . We also found 
that most of the LHLs run East-West connecting locations in the North Atlantic and the US with
the far East. 
We investigated the prevalence of {\em visible} 
MPLS tunnels across LHLs finding an average adoption of values, at the country, continent and AS levels, 
 of 7.79\%, 6.86\%, 11.89\%, respectively, with over 90\% adoption by some operators. 
We found that at a country-level, the US is the preferred long-haul destinations for most regions, 
except for Africa. We closed the section introducing the 
concept of {\em super routers} -- nodes that aggregate multiple LHLs and connect to several 
countries simultaneously --  and showing them to be most commonly found in the US.

%% file: tex/longitudinal.tex
\section{A Longitudinal Perspective}
\label{sec:longitudinal}

In the following paragraphs we explore topological changes in LHL connectivity over a period of 
7 years, starting in 2016. 
To minimize sampling bias introduced by the platform expansion during this 7-year period, we narrow our analysis to the 53 vantage points ($\approx$ \%21.7 of all active probes in that period) in with snapshots in at least 5 years of our dataset.
We start by applying graph-theoretical concepts to investigate changes in the long-haul network ({\em LHnet}) at router, AS and country levels.
We investigate these changes by looking at variations in the node degree distribution (\S\ref{sec:longitudinal:nd_dist}) and in the degree of each node (\S\ref{sec:longitudinal:nd_var}).
We further investigate the changes in the composition of the LHnet over time by looking at the most densely connected nodes (\S\ref{sec:longitudinal:core}) and studying characteristics of the connected components in these graphs (\S\ref{sec:evo:connected_components}).
We conclude our analysis by investigating changes in the per-network inter-hop latency (\S\ref{sec:evo:length}) and the composition of business relationships between networks in both ends of the LHLs (\S\ref{sec:evo:tor}).



\subsection{Changes in the long-haul network}\label{sec:longitudinal:nd_dist}

We investigate changes in the structure of the long-haul network ({\em LHnet}) over time, by looking at it from three different levels,
\ione the router level, \itwo the Autonomous System (AS) level, and \ithree the country level.


%

\begin{figure*}[ht!] \centering 
	\subcaptionbox{\label{fig:dist_deg_evo:router}}{%
		\includegraphics[width=.23\textwidth]{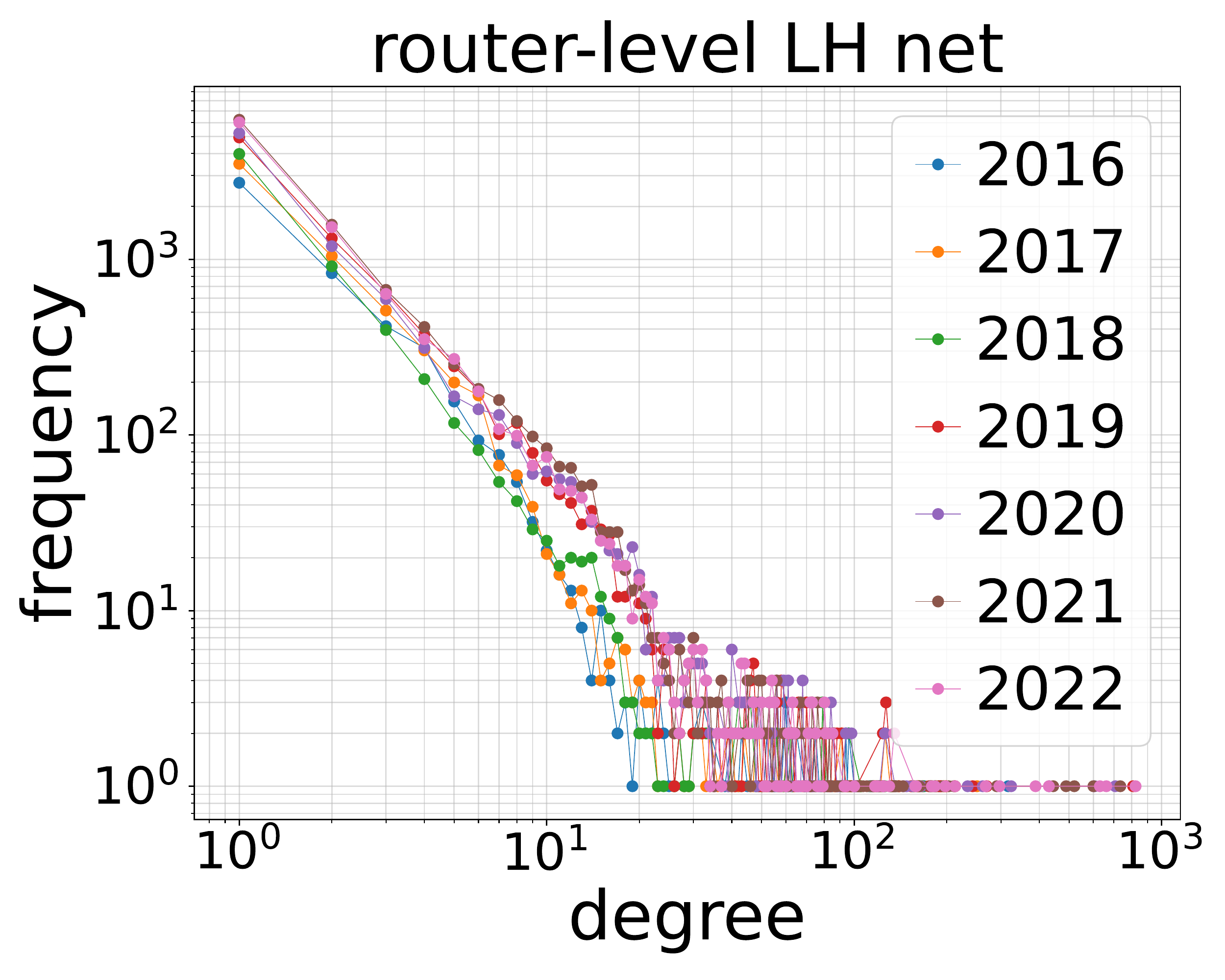}
	}~\subcaptionbox{\label{fig:dist_deg_evo:as}}{%
		\includegraphics[width=.23\textwidth]{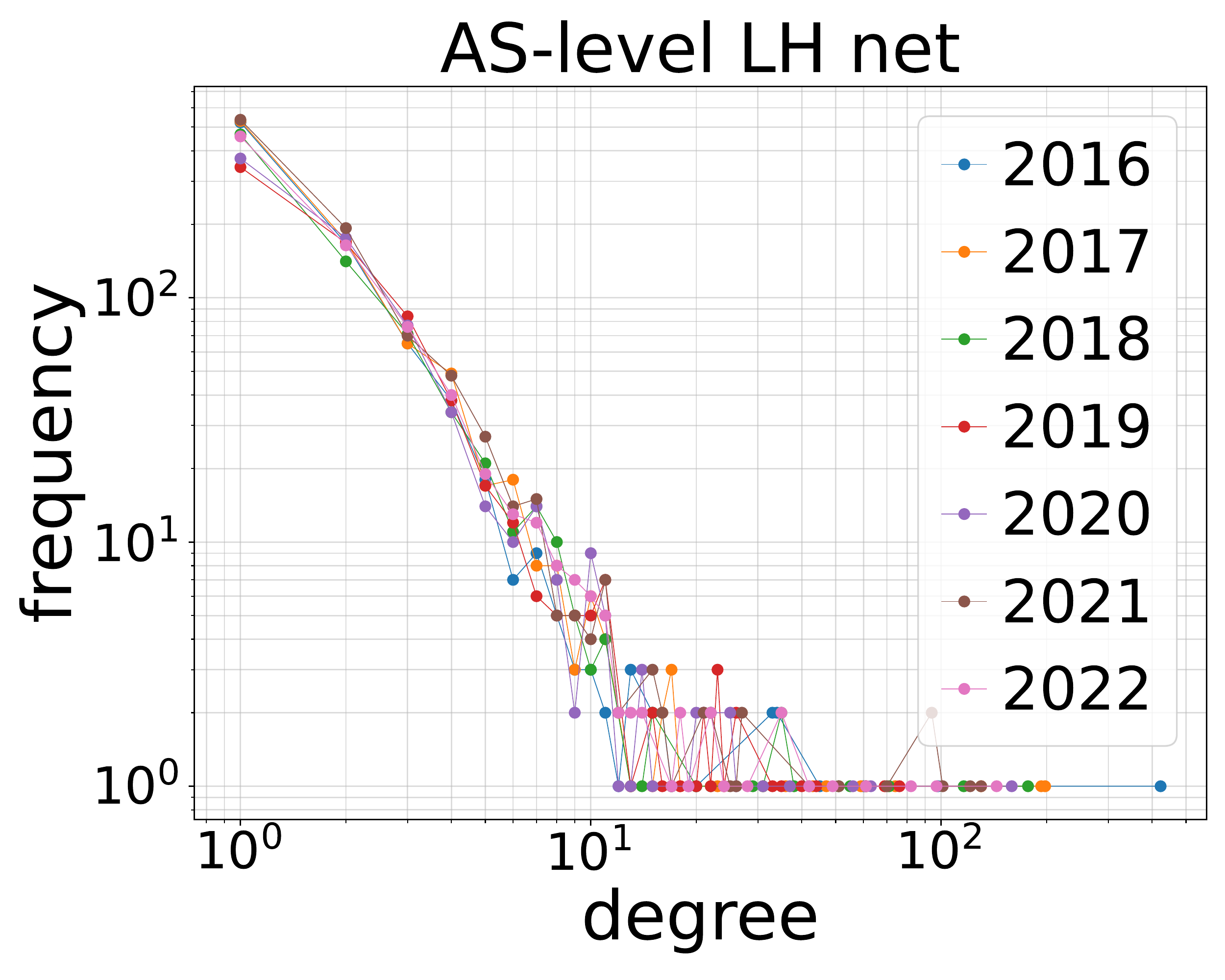}
	}~\subcaptionbox{\label{fig:dist_deg_evo:cc}}{%
		\includegraphics[width=.23\textwidth]{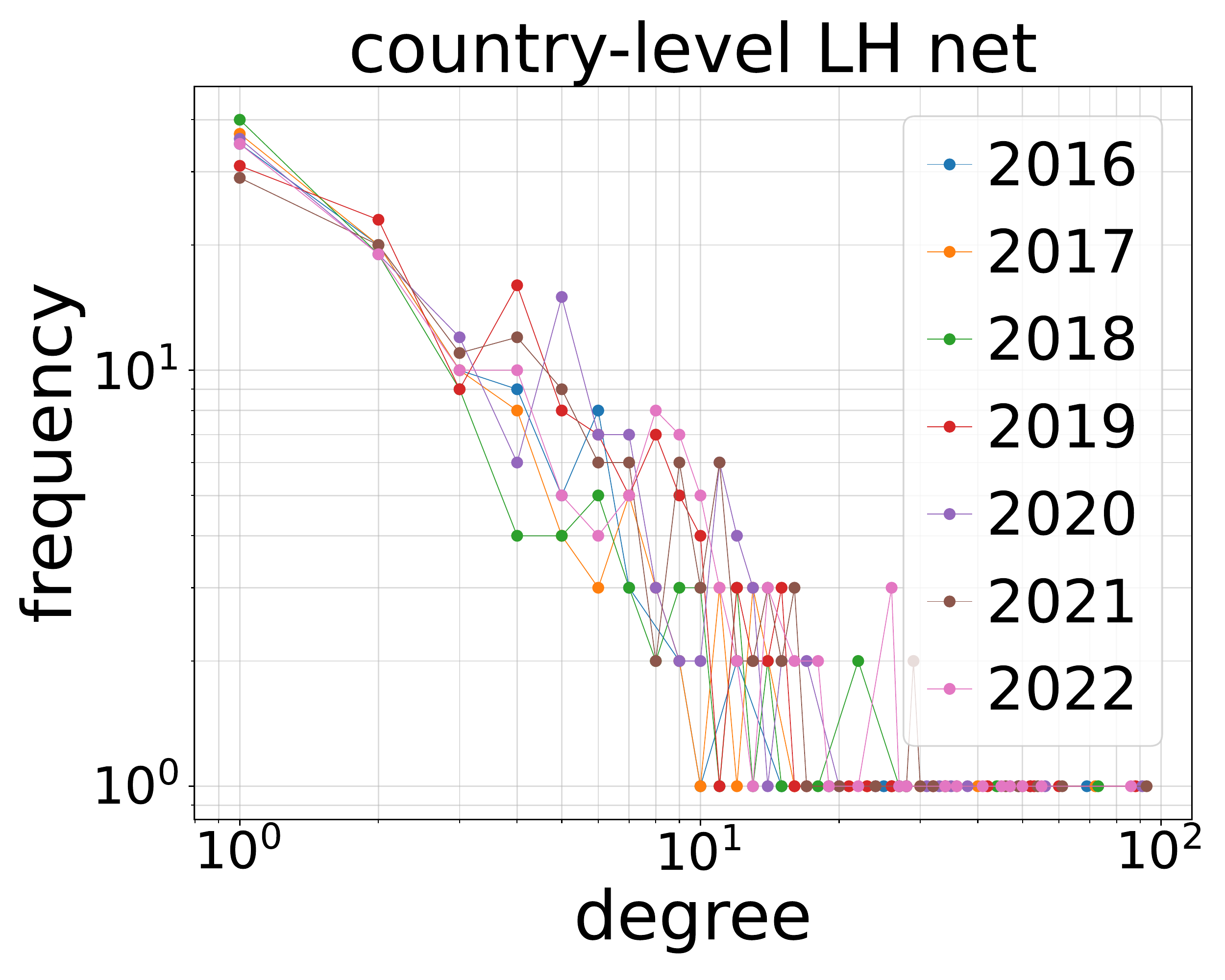}
	}
	\caption{
	Node degree distribution for router, AS and country-level graphs derived from the LHnet.
	} \label{fig:dist_deg_evo}
\end{figure*}




We use a graph to represent the LHnet where nodes represent routers, ASes or countries and edges represent LHLs connecting a pair of nodes.
Figure~\ref{fig:dist_deg_evo} shows the node degree distribution for the router (Fig.~\ref{fig:dist_deg_evo:router}), AS (Fig.~\ref{fig:dist_deg_evo:as}) and country-level (Fig.~\ref{fig:dist_deg_evo:cc}) graphs derived from the LHnet for snapshots over a 7-year period.
In the three cases, the log-log plots show characteristics of {\em heavy-tailed} distributions suggesting that these graphs can be explained by the {\em preferential attachment} model~\cite{albert2002statistical}.
We also compare goodness of fit between log-normal or a powerlaw distributions~\cite{clauset2009power} and find that the powerlaw fits better for AS- and country-level distribution, and at the router level we did not find conclusive evidence for the 7-year dataset.
We also fit powerlaw to each node degree distribution to obtain the characteristic parameter and find minor fluctuation over time but a subtle shift towards a slower tail decay.


\subsection{Node degree changes over time}\label{sec:longitudinal:nd_var}

Given that the our graph-theoretical observations show changes in the structure of the network over time, we further investigate individual changes in the connectivity of nodes in the AS- and country-level graphs.

\begin{figure}[httb!] \centering 
	\subcaptionbox{\label{fig:var:as}}{%
		\includegraphics[width=.23\textwidth]{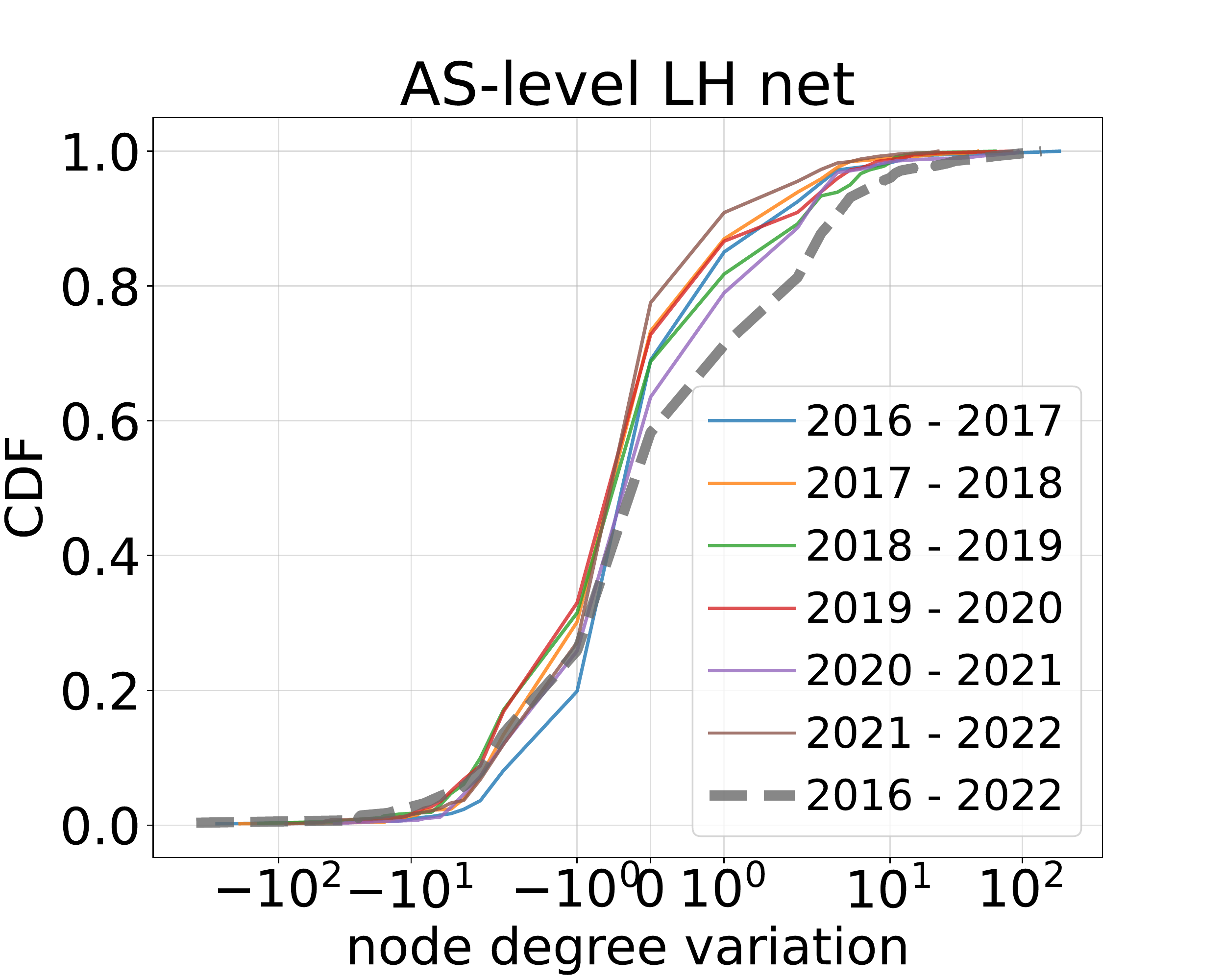}
	}~\subcaptionbox{\label{fig:var:cc}}{%
		\includegraphics[width=.23\textwidth]{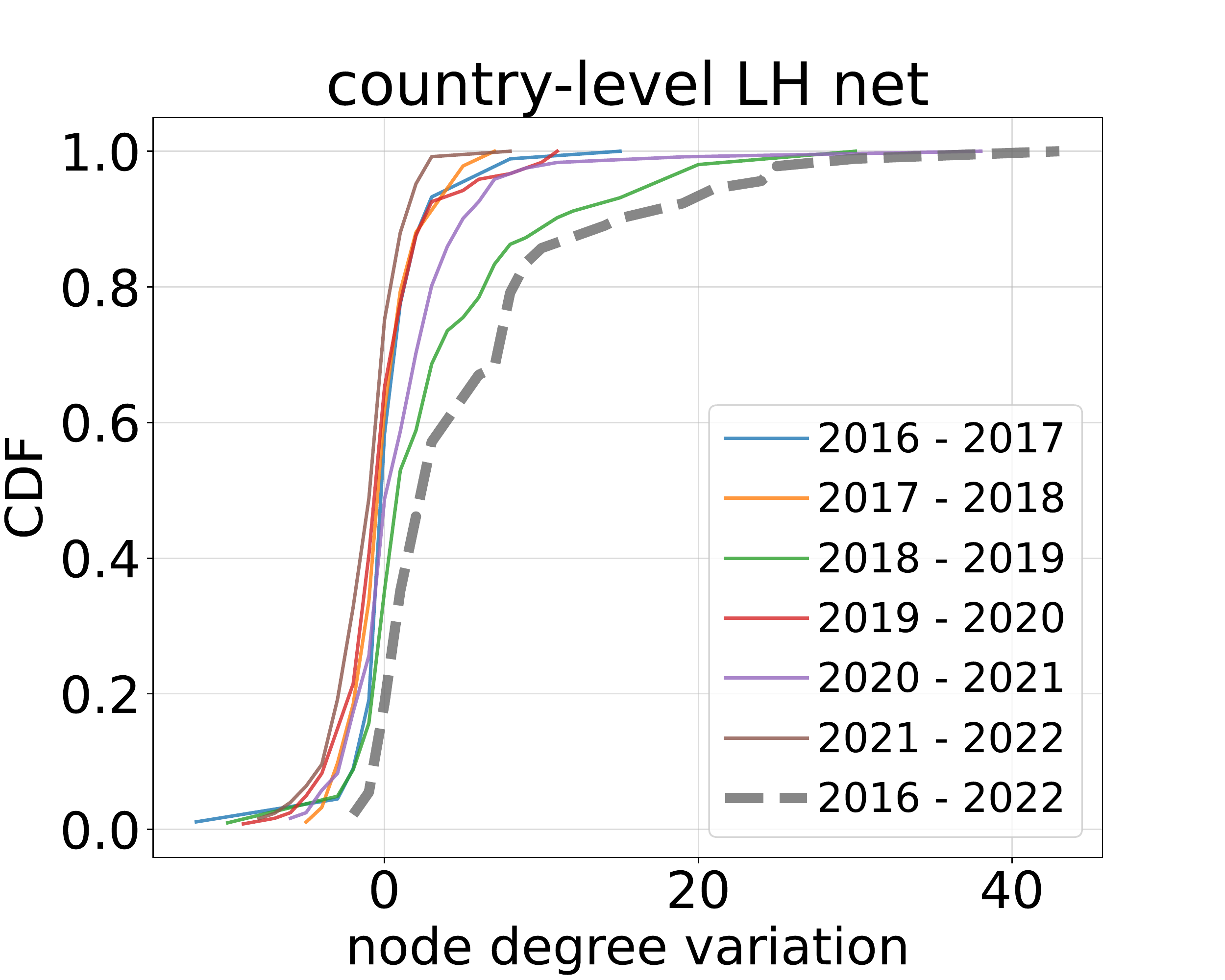}
	}
	\caption{
		Distribution of YoY and 2016-2022 variations of node degrees in the  AS and country-level graphs.
	} \label{fig:var}\vspace{-10pt}
\end{figure}

Figure~\ref{fig:var} shows the cumulative distribution of the year-to-year and 2016-to-2022 variations in the node degree of nodes in the AS- (Fig.~\ref{fig:var:as}) and country-level (Fig.~\ref{fig:var:cc}) graphs.
We observe that both graph have major changes over time, at the AS-level, the  2016-to-2022 distribution shows a symmetric distribution meaning neutral changes overall, while at the country-level is skewed towards positive values with a mean node degree variation of 5.79.
These changes indicate that the LHnet has been reshaped at the AS-level but new nodes replace connectivity of declining nodes, on the other hand, changes at the country level show a that countries are getting more densely connected over time.

\subsection{A stable LHnet core}\label{sec:longitudinal:core}

We now extend our graph-theoretical analysis and investigate the prevalence over time of the group of most connected nodes.

%



\begin{table}[ht!]
\centering\footnotesize
\begin{tabular}{r|p{5cm}}
\toprule
		& TOPcore members \\
\hline
AS level	& Cogent-174, HE-6939, Arelion-1299, LUMEN-3356, Telecom Italia-6762, TATA-6453, PNG-9299 \\
\hline
Country level	& South Africa, India, Germany, France, Great Britain, United States, Hong Kong, the Netherlands, Singapore, Australia, Italy, Turkey, Canada and Brazil\\
\bottomrule
\end{tabular}

\caption{TOPcore members at AS and county levels present at least in 6 different snapshots}\label{table:topcore}

\end{table}

We apply {\em k}-core decomposition to generate discrete tiers of nodes~\cite{JIAH2008, carisimo2019studying}, where the presence in each tier is given with the connectivity of a node and its neighbors.
We then identify the TOPcore (tier with the highest shell-index) and look for ASes and countries with long-term presence at the TOPcore.
Table~\ref{table:topcore} shows ASes and countries that were in the TOPcore in 6 different years.
The AS-level graph TOPcore has a stable group os ASes composed of are large international carriers (\eg Cogent-AS174, Hurricane Electric-AS6939, LUMEN-AS3356) that are typically inferred to be part of the {\em transit-free clique}~\cite{luckie2013relationships}.
At a country level, the stable TOPcore is composed of countries in all continents, though a larger prevalence in the North Atlantic.
These countries are known for being international or regional hubs, with vast presence of submarine cable networks (\eg the US, Brazil, India and Singapore) or hosting large IXPs (DECIX, AMS-IX, IX.br).

\subsection{The long-haul network}
\label{sec:evo:connected_components}


We continue the longitudinal graph-theoretical analysis of the LHnet by looking at the evolution of  \textit{connected components} in the router-, AS- and country-level graphs.

\begin{table*}
\centering
\footnotesize

\begin{tabular}{c|ccc|cc|ccc|cc|ccc|cc}
\toprule
&	\multicolumn{5}{c|}{router-level graph} &	\multicolumn{5}{c|}{AS-level graph} &	\multicolumn{5}{c}{country-level graph} \\
\hline
		&	\multicolumn{3}{c|}{LHnet ($G$)}		& \multicolumn{2}{c|}{$\max(C_G)$} &	\multicolumn{3}{c|}{LHnet ($G$)}		& \multicolumn{2}{c|}{$\max(C_G)$} &	\multicolumn{3}{c|}{LHnet ($G$)}		& \multicolumn{2}{c}{$\max(C_G)$}\\
cycle		& |$C_G$|&	|$V$|	& $|E|$ & 	$|V|$	& $|E|$ & $|C_G|$&	$|V|$	& $|E|$ & 	$|V|$	& $|E|$ & $|C_G|$&	$|V|$	& $|E|$ & 	$|V|$	& $|E|$  \\
\hline
\hline

2016	&521	&4863	&7857	&2813	&5953	&24	&777	&1002	&727	&975	&1	&102	&260	&102	&260\\
2017	&634	&6069	&9516	&3122	&6522	&22	&808	&1092	&760	&1065	&1	&107	&291	&107	&291\\
2018	&721	&6033	&8285	&2773	&5439	&26	&724	&955	&666	&922	&1	&107	&301	&107	&301\\
2019	&904	&8465	&14888	&3211	&9512	&32	&582	&783	&499	&731	&1	&139	&532	&139	&532\\
2020	&832	&8467	&15659	&3944	&10335	&42	&627	&824	&538	&776	&1	&139	&536	&139	&536\\
2021	&894	&10336	&18942	&6382	&15262	&32	&837	&1146	&764	&1104	&2	&139	&630	&137	&629\\
2022	&903	&9802	&17224	&2880	&9075	&35	&712	&950	&621	&893	&2	&136	&576	&134	&575\\
\bottomrule
\end{tabular}

\caption{
Graph dimension of the {\em LHnet} and the largest connected component. 
} \label{table:connected_components}

\end{table*}

%
%
%
%

Table~\ref{table:connected_components} shows the evolution of the number of vertices, edges and connected components and the size of the sub-graph containing the largest connected component (nodes and vertices) for the router-, AS- and country-level graphs in a 7-year timeframe.
Notably, the three graphs show that the largest connected component comprises a large fraction (or even all) nodes creating a contiguous network that represents {\em the intercontinental long-haul backbone of the public Internet}. 
At level that, despite the LHnet being composed of hundreds of connected components, the largest connected component includes between 29.3\% to 61.7\% of nodes and 52.6\% to 80.5\% of links.
For AS- and country-level perspective, the connected component comprises up to 93.5\% and 97.3\% of the nodes and 99\% and 100\% of the edges, respectively.
These observations indicate links and routers visible in traceroute campaigns create a contiguos intercontinental network containing 621 ASes ($\approx$ 1\% of all active ASes in October 2022) and 134 countries ($\approx$ 70\% of all countries).

\subsection{The LHL length variations}
\label{sec:evo:length}

Based on our previous observations that showed some changes in the characteristics of the LHnet, we now focus on the inter-router latency.
We use our multi-year dataset to investigate changes in the AS-level inter-router latency that could describe modifications in the structure of the network that were not visible from previous macroscopic granularities.

\begin{figure}[th!] 
	\centering 
	\includegraphics[width=.36\textwidth]{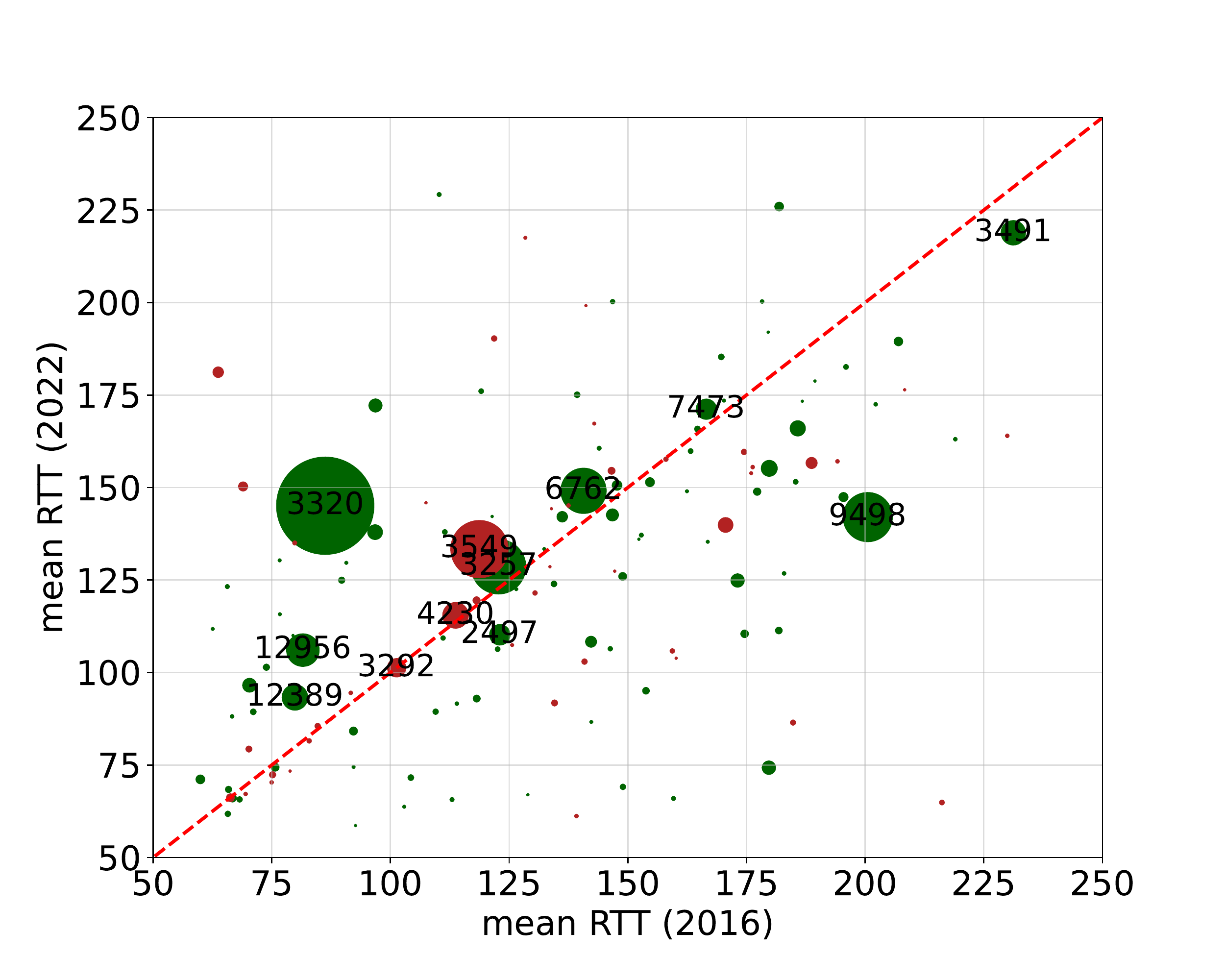}
	\caption{
		Variations in the mean LHL inter-router RTT at the AS level.
		Sizes and colors show per-AS link variations and sign.
	}
	\label{fig:length_var}
\end{figure}

To investigate changes in the composition of long-haul connectivity at AS level, we compare the mean LHL inter-router latency for networks present in the 2016 and 2022 snapshots.
Figure~\ref{fig:length_var} shows a scatter plot that axis indicate the mean LHL inter-router latency in the 2016 and 2022 snapshots and the size of the dots is a function of the variation in the number of per-AS LHL where colors indicate positive (green) or negative (red) variations.
In this figure, the red-dashed diagonal line indicate whether networks increased or decreased their mean inter-router latency if their dots are above or below the line.
We focus on networks with prominent variations in the number of LHLs ($>$100) where we observe a wide variety of trends in this 6-year period.
There is a group of large providers with neutral changes in the mean LHL inter-router latency despite having a growth (Telecom Italia-6762, SingTel-7473, GTT-3257, IIJ-2497) or a decay (Level3-3549, Claro Brasil-4230, TDC-3292) in the number of LHLs.
Another example of stability over time is PCCW with the largest mean LHL inter-router latency (despite a significant growth) in our analysis which is given by a large fraction of LHLs connecting far distant locations such as Miami and Frankfurt or Tokyo and Frankfurt.
This figure show some networks with clear changes in the number of LHLs and the mean LHL inter-router latency, DTAC-3320 (+3214 LHLs, +58.2ms), Telxius-12956 (+348 LHLs, +24.5ms), Bharti Airtel-9848 (+808 LHLs, -58.5ms).
Several reasons can explain these changes such as deployments of new submarine cables, organization-level reconfigurations (Level3), mergers and acquisitions (DTAC purchase of Sprint~\cite{tmobile:sprint}), and technological upgrades (adoption of MPLS).

%
%
%



\subsection{Commercial relationships behind LHLs}
\label{sec:evo:tor}

We shift our attention to the AS relationship between both ends of long-haul links. For this 
part of the analysis we use CAIDA's AS relationship files~\cite{asrel, giotsas2014inferring}, 
curated from both BGP and traceroute-derived sources from RouteViews and RIPE RIS collectors.

\begin{table}[hbt!]

	\centering\footnotesize

\caption{Intra-/inter-domain LHLs (and unmapped) over time. 
} \label{table:tor1}

	\setlength\tabcolsep{1.5pt}
	\begin{tabular}{c c c c c}
		\toprule
		Year    & Intra-domain	& Inter-domain & Unmapped	&\# LHL \\
		\hline
		\hline
2016	&5637 (0.72)	&2207 (0.28)	&13 (0.00)	&7857\\
2017	&7231 (0.76)	&2247 (0.24)	&38 (0.00)	&9516\\
2018	&6228 (0.75)	&2057 (0.25)	&0 (0.00)	&8285\\
2019	&12496 (0.84)	&2254 (0.15)	&138 (0.01)	&14888\\
2020	&13808 (0.88)	&1849 (0.12)	&2 (0.00)	&15659\\
2021	&16268 (0.86)	&2674 (0.14)	&0 (0.00)	&18942\\
2022	&14954 (0.87)	&2270 (0.13)	&0 (0.00)	&17224\\
		\bottomrule
	\end{tabular}

\end{table}

We first look at the fraction of long-haul links that are intra- or inter-domain links. 
Table~\ref{table:tor1} list both, over time, and their respective fractions.  
We observe a rapid growth in the number of LHLs -- which doubles over the observation period -- with the fractions of intra-domain LHLs growing at fastest pace.
Throughout the period of analysis, the vast majority LHLs in our dataset are intra-domain (between 77 and 88\%). 

For the remaining LHLs, we list the fraction of links based on the categories that can be inferred 
in Table~\ref{table:tor}. We find that roughly one-fifth of the LHLs correspond to inter-domain links, with 
customer-to-provider (c2p) links responsible for two-thirds of the inter-domain LHLs. In c2p and p2p scenarios, 
it is difficult to identify which of the two sides provides the physical connectivity by either allocating its 
own resources or purchasing capacity. In some cases, we can not find any inferred AS relationship between 
reported ASes at either end of the LHL. A small fraction of these non-inferred relationships 
corresponds to incorrectly mapped peering relationships at IXPs. 

\begin{table}[hbt!]
\centering

\caption{Commercial relationships between ASes at both ends of Inter-domain LHLs. 
	} \label{table:tor}

\setlength\tabcolsep{1.5pt}
\footnotesize
\begin{tabular}{c c c c c}
\toprule
year & \multicolumn{2}{c}{Inter-domain} & \multicolumn{2}{c}{non-inferred ToR}\\
     & p2p	&	p2c & IXP	& Unknown \\
\hline
\hline
4577	&665 (0.30)	&1112 (0.50)	&0 (0.00)&430 (0.19)	\\
5423	&475 (0.21)	&1224 (0.54)	&0 (0.00)&548 (0.24)	\\
6447	&462 (0.22)	&1074 (0.52)	&0 (0.00)&521 (0.25)	\\
7616	&681 (0.28)	&1251 (0.52)	&0 (0.00)&322 (0.13)	\\
8821	&420 (0.23)	&1127 (0.61)	&0 (0.00)&302 (0.16)	\\
9644	&521 (0.19)	&1713 (0.64)	&0 (0.00)&440 (0.16)	\\
10020	&389 (0.17)	&1363 (0.60)	&0 (0.00)&518 (0.23)	\\
\bottomrule
\end{tabular}
\end{table}

\subsection{Takeaways}

In this section we explored the evolution long-haul connectivity over time.
We found minor fluctuations in node degree distribution at the router, AS and country levels, though, with changes towards powerlaw distributions with lower $\alpha$ values.
At a country level, we also discovered that the network is more densely connected meaning that new LHLs were created interconnecting new pairs of countries.
We showed that the LHnet is composed by a stable group of networks and countries in its core and the largest connected component of the long-haul 
network contains up to 62\% of all nodes linked by LHLs making it {\em the intercontinental long-haul backbone 
of the public Internet}.
We also found a wide range of changes in the inter-router latency at the AS level, with some networks creating new and longer LHLs.
%


%% file: tex/implications.tex
\section{Practical implications and limitations}
\label{sec:implications}

In this section, we discuss practical implications of our observations and briefly touch on limitations of our study.
The network-level perspective of the intercontinental long-haul infrastructure offers a new perspective on
the structure of the network connectivity, and raises questions on a wide-range of topics, from network management
and operations, to congestion control algorithms, network resilience and cyber sovereignty.

{\bf Management and operations}: 
The opaqueness of a tunneled network structure reduces the ability to debug the network through ICMP-based tools (\ie pings and traceroutes).
The lack of basic path-discovery tools capable of mapping elements within tunnels impedes the ability to easily detect path changes caused by traffic engineering (\eg load balance) or reconfigurations (\eg rerouting, outages).
In absence of these tools, debugging is only available for a group of network operators with privileged access to devices in the network (\eg MPLS switches) using sophisticated tools.

{\bf Traffic engineering}:
A previous study conducted in Microsoft's WAN network showed persistent latency inflation caused by MPLS {\em autobandwidth} algorithms~\cite{pathak2011latency}.
Other research studies observed path inflation and speculate that presence of MPLS tunnels could explain this behavior~\cite{bozkurt2018dissecting, krishnan2009moving}.
In the context where LHLs are the result of MPLS tunnels, they could experience similar path and latency inflation.

\textbf{Long-haul resilience.} The change towards a network core with fewer routers, enabled 
by the growing adoption of link-layer technologies like MPLS, brings up resilience concerns 
already argued by Vanaubel et al.~\cite{vanaubel2017through} to which we add the role and criticality 
of {\em super routers}. Technologies such MPLS~\cite{vanaubel2017through} or SD-WAN~\cite{jain2013b4, 
hong2013achieving}, increase the opaqueness of the network, limits our understanding of underlying 
physical characteristics of the network, and challenges a thorough assessment of network resilience.

\textbf{LHL and submarine cables.}
Private conversations with operators affiliated to the SubOptic community, confirm the adoption of
network virtualization and other techniques to \textit{move landing points inland} 
to the proximity of datacenters, suggesting that the list of threats to submarine 
cable connectivity should include those of the infrastructure connecting the cable's 
landing point with its final destination. This is well illustrated by the city-wide 
network outage that the Colombian city of Cali experienced in 2021 (visible from IODA~\cite{ioda-co}) due 
to a cable cut in the 100km cable segment that connects the city with the submarine 
landing point~\cite{co2021outage}.


{\bf Cyber sovereignty}:
In recent years we observed governments' concerns about cyber sovereignty bringing consequent updates to regulatory frameworks, such as GDPR~\cite{gdpr} and LPDG~\cite{lgpd}.
These concerns sometimes include the routing system which may be also subject to regulations~\cite{obar2012internet, hathaway2014connected}.
Following those steps, research efforts evaluated whether end-to-end paths applied forwarding rules skipping specific countries. 
The vast adoption of tunneling mechanisms, however, challenges such assessments and attempts to validate regulation compliance 
or precisely identify potential vulnerable choke points.

{\bf Climate change threats}:
Consequences of sea-level rise for coastal network infrastructure is a growing concern across multiple entities including the UN~\cite{un:climate}, 
researchers~\cite{npr:climate}, 
the submarine cable organizations~\cite{icpc:climate} and other communities.
Being unable to identify physical infrastructure underneath these network-layer links limits our ability to identify LHL exposed to natural disasters and climate change threats.

\textbf{Limitations.} We acknowledge several limitations of this study, to help put our 
findings in perspective. For starters, the presence of routers that ignore ICMP messages and 
middleboxes~\cite{Sherry:MB} (\eg NAT, firewalls, \etc), may impact our latency-derived 
estimations. We believe, however, that our main observations such as on the 
scale and rapid growth of the long-haul infrastructure should hold. Our reliance on 
measurement collected from a volunteers platform may introduce biases on our observations resulting 
from its uneven adoption, while inaccuracies in topological datasets (\eg geolocation databases, alias 
resolution, router-to-AS mappings) may affect the topologies we generate. 
Our primary approach for addressing these issues was to be conservative in all of our design and implementation choices to the limit the impact on our results.

 Appendix~\ref{sec:appendix:sensitivity} includes reults from a sensitivity analysis including timeframe, dataset size, threshold variations and filters' contributions.

%% file: tex/relwork.tex
\section{Related Work}
\label{sec:relwork}

The router- and AS-level Internet topologies have been widely studied from multiple angles, including graph-theoretical models~\cite{Faloutsos:powerlaw, PhysRevLett.85.4626}, commercial relationships~\cite{gao2001stable, luckie2013relationships}, flattening and rewiring~\cite{dhamdhere2010internet}, among others.

Several research efforts have focused on documenting structural changes of the Internet to accommodate the rise of new technologies (\eg video streaming, smartphones).
The transition from a hierarchical network into a flat structure~\cite{dhamdhere2010internet} in the early 2000s is well documented.
The irruption of CDNs gained a large deal of attention with studies focusing on changes in topological and traffic characteristics~\cite{labovitz2010internet}, the widespread of direct peer-to-peer connectivity of CDNs~\cite{chiu2015we}, {\em off-net} cache deployments~\cite{gigis2021seven}.
This transition also included the consolidation of IXPs as key pieces of the Internet topology, creating peering fabrics with levels traffic similar to Tier-1 Transit providers~ \cite{ager2012anatomy}, and expanding to all regions~\cite{fanou2015diversity, carisimo2020first}.

A handful of previous studies have focused on long-haul connectivity. Durairajan {\em et al.}~\cite{ram:intertubes} investigated the domestic long-haul infrastructure of the United States using information extracted from optical cable deployments along pre-existing transport infrastructure. Bishof et al.~\cite{bischof2018untangling} puts forward a research agenda focused on the criticality of the submarine cable network. Fanou {\em et al.}~\cite{fanou2020unintended} studied the impact of new submarine cable deployments in developing regions and the drop in latency to cross the Atlantic. Liu {\em et al.}~\cite{liu2020out} found that submarine cable infrastructure enables fetching resources contained in most popular websites.

International connectivity is at times closely related to geopolitics and foreign affairs. Levin {\em et al.}~\cite{10.1145/2785956.2787509} investigated traffic censorship of intermediary countries along the path of international routes and routing avoidance of specific censorship regions. 
Future research directions could combine long-haul connectivity and its implications with geopolitics.

%% file: tex/conclusions.tex
\section{Conclusions and Future Work}

This study contributes a new perspective of Internet topology focused on the intercontinental 
long-haul connectivity. We presented a methodology for identifying long haul links (LHLs) 
in traceroute measurements, and reported on our analysis of the long-haul infrastructure 
using a large corpus of traceroute data collected at the edge of the network. We found 
a vast and rapidly growing network with links spanning over 10,000km and nodes that connect 
as many as 45 countries. Despite its rapid growth, we found a graph with key characteristics
and a core component that remain stable over time. This new perspective opens a wide range of
promising directions for future research, from alternative views of the long haul infrastructure
to an exploration of that infrastructure's key properties and temporal stability. 

%% file: tex/appendix.tex
\section{Ethics}

This work does not raise any ethical issues.

%
%
%
%
\lstdefinestyle{myCustomMatlabStyle}{
  keywords={\[x\],},
alsoletter={\[ \]},
morekeywords={},
keywordstyle=\color{blue}\bfseries,
aboveskip=20pt,
belowskip=20pt,
identifierstyle=\color{black},
morecomment={[n][\color{purple}]{\#}{\^^M}},
numbers=left,
numberstyle=\color{black}\scriptsize,
rulecolor=\color{black},
stepnumber=1,
numbersep=8pt,
showstringspaces=false,
breaklines=true,
postbreak=\noindent,
breakautoindent=false,
frame=single,
backgroundcolor=\color{background},
}
\lstset{basicstyle=\scriptsize\ttfamily,style=myCustomMatlabStyle}
\section{Examples of traceroute measurements traversing a {\em super router}}\label{sec:appendix:tr:superrouters}

Listings~\ref{lst:telia:1} and \ref{lst:telia:2} show traceroute measurements collected by CAIDA's Ark monitors {\tt jfk-us} (Lst.~\ref{lst:telia:1}) and {\tt ord-us} (Lst.~\ref{lst:telia:2}) traversing Telia's (AS1299) {\em super router} in Chicago during the measurement cycle 8820 in October 2020.
These results show that the Chicago router is the ingress point to a global link-layer backbone with egress points in different major cities across the United States (Los Angeles and Seattle) and the world (Budapest and Finland).

\begin{lstlisting}[
caption={Traceroute \#1 traversing Telia's super-router in Chicago. FROM jfk-us (jfk-us.team-probing.c008820.20201002.warts.gz)},
label={lst:telia:1},
literate=
*{:}{{{\color{black}{:}}}}{1}
{\{}{{{\color{black}{\{}}}}{1}
{\}}{{{\color{black}{\}}}}}{1}
{,}{{{\color{black}{,}}}}{1},
captionpos=b]
# traceroute from 216.66.30.102 (Ark probe hosted in New York City, NY, US. No AS info found) to 223.114.235.32 (MAXMIXD: Turpan, CN)
 1  216.66.30.101	0.365 ms
 2  62.115.49.173	3.182 ms
 3  *
 4  62.115.137.59	17.453 ms [x] 	(chi-b23-link.ip.twelve99.net., CAIDA-GEOLOC -> Chicago, IL, US)
 5  62.115.117.48	59.921 ms [x] 	(sea-b2-link.ip.twelve99.net., RIPE-IPMAP -> Seattle, WA, US)
 6  62.115.171.221 	69.993 ms
 7  223.120.6.53	69.378 ms
 8  223.120.12.34	226.225 ms
 9  221.183.55.110	237.475 ms
10  221.183.25.201	238.697 ms
11  221.176.16.213	242.296 ms
12  221.183.36.62	352.695 ms
13  221.183.39.2	300.166 ms
14  117.191.8.118	316.270 ms
15  *
16  *
17  *
18  *
19  *
\end{lstlisting}

\begin{lstlisting}[
caption={Traceroute \#2 traversing Telia's super-router in Chicago. FROM ord-us (ord-us.team-probing.c008820.20201002.warts.gz)},
label={lst:telia:2},
literate=
*{:}{{{\color{black}{:}}}}{1}
{\{}{{{\color{black}{\{}}}}{1}
{\}}{{{\color{black}{\}}}}}{1}
{,}{{{\color{black}{,}}}}{1},
captionpos=b]
# traceroute from 140.192.218.138 (Ark probe hosted in Chicago, IL, US at Depaul University-AS20130) to 109.25.215.237 (237.215.25.109.rev.sfr.net., MAXMIXD: La Crau, FR)
 1  140.192.218.129  0.795 ms
 2  140.192.9.124  0.603 ms
 3  64.124.44.158  1.099 ms
 4  64.125.31.172  3.047 ms
 5  *
 6  64.125.15.65  1.895 ms      [x] (zayo.telia.ter1.ord7.us.zip.zayo.com., CAIDA-GEOLOC -> Chicago, IL, US)
 7  62.115.118.59  99.242 ms    [x] (prs-b3-link.ip.twelve99.net., CAIDA-GEOLOC -> Paris, FR)
 8  62.115.154.23  105.214 ms
 9  77.136.10.6  119.021 ms
10  77.136.10.6  118.830 ms
11  80.118.89.202  118.690 ms
12  80.118.89.234  118.986 ms
13  109.24.108.66  119.159 ms
14  109.25.215.237  126.085 ms
\end{lstlisting}

\section{Sensitivity Analysis}\label{sec:appendix:sensitivity}

We run a preliminary analysis to investigate the sensitivity of our observations to timeframe, dataset size and threshold variations. 
We also analyze the results obtained by removing geolocation and latency filters from our methodology (\S\ref{sec:method}).

To address in part some of these limitations, 
we ran a simple analysis of the sensitivity of our findings to timeframe, dataset size and threshold variations.
We analyzed fluctuations
over a one-week period --- a 
timeframe in which the network is expected to be stable --- finding no changes in the characteristics of the 
long-haul connectivity. We also evaluate the impact of the dataset size in our observations finding that a 
downsample of 1:2 and 1:4 still captures a fraction of 0.97 and 0.82 of the nodes and 0.96 and 0.76 of the links, 
respectively. Our evaluation of the sensitivity to changes of the LHL threshold varying it from 57ms to 20ms found only a 3.4\% variation on the number of nodes and a 2.4\% variations in the number of links.


\begin{figure}[th!] 
	\centering 
	\includegraphics[width=.4\textwidth]{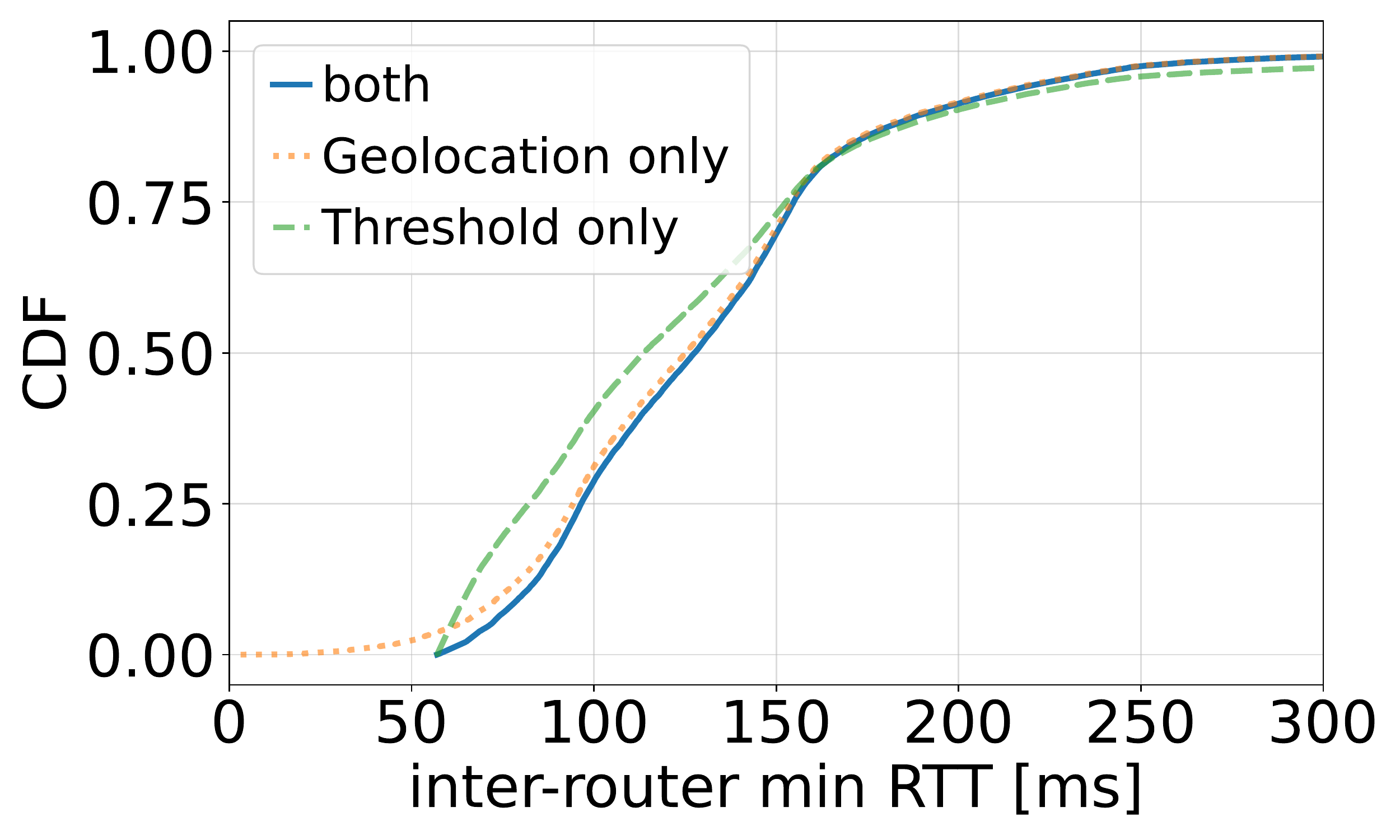}
	\caption{
		Cumulative distribution of long-haul inter-router latency. $\approx$75\% of the intercontinental inter-router latency difference ranges between 60 and 155ms.
		}
	\label{fig:lhl_rtts:all}
\end{figure}

We use our results to compare the output of the methodology (\S\ref{sec:method}) if only geolocation-based or latency-based filters were applied.
Figure~\ref{fig:lhl_rtts:all} is a CDF of inter-router latency differences (in milliseconds) for our methodology (combines geolocation data with the LHL threshold) and the output of geolocation-based or latency-based alternatives.
Despite not being identical, we observe that the three curves have similar profiles with Jensen-Shannon scores of 0.10 and 0.15 between our method and geolocation-based and latency-based, respectively.
This similarity in the results obtained with the three alternatives suggests that the implications discussed for intercontinental LHLs are going to be shared by other links of the same range.